\documentclass[aps,prd,twocolumn,groupedaddress,noeprint,longbibliography,superscriptaddress]{revtex4-1}

\usepackage{color}
\usepackage[normalem]{ulem}
\usepackage{enumitem}
\usepackage{verbatim}
\usepackage{afterpage}

\usepackage{amsmath, amssymb}
\usepackage{mathtools}

\allowdisplaybreaks
\usepackage{graphicx}

\usepackage{braket}

\newcommand{\para}[1]{\left( #1 \right)}

\makeatletter
\renewcommand{\env@cases}[1][@{}l@{\quad}l@{}]{%
	\let\@ifnextchar\new@ifnextchar
	\left\lbrace
	\def\arraystretch{1.2}%
	\array{#1}%
}
\usepackage{tensor}
\newcommand{\idx}{\indices}
\newcommand{\de}{\partial}
\newcommand{\di}[1][]{\mathrm{d}^{#1}}
\newcommand\numberthis{\addtocounter{equation}{1}\tag{\theequation}}
\usepackage{kbordermatrix}
\renewcommand{\kbldelim}{(}
\renewcommand{\kbrdelim}{)}

\usepackage[utf8]{inputenc}
\usepackage{amsthm}

\usepackage{array}
\usepackage{hhline}
\usepackage{tabularx}
\newcolumntype{Y}{>{\centering\arraybackslash}X}
\usepackage{longtable}
\usepackage{makecell}
\usepackage[usestackEOL]{stackengine} 

\usepackage{cleveref} 
\crefname{appch}{Appendix}{Appendices}

\newcounter{magicrownumbers}
\newcommand{\rownumber}{\refstepcounter{magicrownumbers}\arabic{magicrownumbers}}
\newcommand{\refrownumber}[1]{\refstepcounter{magicrownumbers}\label{#1}\arabic{magicrownumbers}}

\usepackage{ifdraft}

\ifdraft{
	\usepackage[colorinlistoftodos]{todonotes}
	
}{
	\usepackage[colorinlistoftodos]{todonotes}
	
}

\AtBeginDocument{%
	\newwrite\bibnotes
	\def\bibnotesext{Notes.bib}
	\immediate\openout\bibnotes=\jobname\bibnotesext
	\immediate\write\bibnotes{@CONTROL{REVTEX41Control}}
	\immediate\write\bibnotes{@CONTROL{%
			apsrev41Control,author="08",editor="1",pages="0",title="0",year="1"}}
	\if@filesw
	\immediate\write\@auxout{\string\citation{apsrev41Control}}%
	\fi
}%

\usepackage{balance}
\usepackage{lastpage}

\AtBeginEnvironment{thebibliography}{\balance}

\begin{document}


\title{Ghost and tachyon free Poincar\'e gauge theories: A systematic approach}


\author{Yun-Cherng~\surname{Lin}}
\email{ycl54@mrao.cam.ac.uk}
\affiliation{Astrophysics Group, Cavendish Laboratory, JJ Thomson Avenue,
Cambridge CB3 0HE, United Kingdom}
\affiliation{Kavli Institute for Cosmology, Madingley Road, Cambridge CB3 0HA, United Kingdom}
\author{Michael P.~\surname{Hobson}}
\email{mph@mrao.cam.ac.uk}
\affiliation{Astrophysics Group, Cavendish Laboratory, JJ Thomson Avenue,
Cambridge CB3 0HE, United Kingdom}
\author{Anthony N.~\surname{Lasenby}}
\email{a.n.lasenby@mrao.cam.ac.uk}
\affiliation{Astrophysics Group, Cavendish Laboratory, JJ Thomson Avenue,
Cambridge CB3 0HE, United Kingdom}
\affiliation{Kavli Institute for Cosmology, Madingley Road, Cambridge 
CB3 0HA, United Kingdom}


\date{\today}

\begin{abstract}
A systematic method is presented for determining the conditions on the
parameters in the action of a parity-preserving gauge theory of
gravity for it to contain no ghost or tachyon particles. The technique
naturally accommodates critical cases in which the parameter values
lead to additional gauge invariances. The method is implemented as a
computer program, and is used here to investigate the particle content
of parity-conserving Poincar\'e gauge theory, which we compare with previous
results in the literature. We find 450 critical cases that are free of
ghosts and tachyons, and we further identify 10 of these that are also
power-counting renormalizable, of which four have only massless
tordion propagating particles and the remaining six have only a massive
tordion propagating mode.
\end{abstract}

\pacs{Add PACS}

\maketitle

\section{Introduction}

Following the gauging of the Lorentz group by Utiyama
\cite{Utiyama1956}, Kibble was the first to gauge the Poincar\'e group
\cite{Kibble1961}. In Kibble's model, the gauge fields of the
Poincar\'e group are $h\idx{_A^\mu}$ and $A\idx{^{AB}_\mu}$, which
correspond to translations and Lorentz transformations, respectively.
Such Poincar\'e gauge theories (PGTs) have a geometric interpretation
in terms of a Riemann--Cartan spacetime ($U_4$), which differs from
the more familiar Riemann spacetime ($V_4$) in having nonzero
torsion. In this geometric interpretation, the field strengths of the
translational and rotational gauge fields are identified as the
torsion and curvature, respectively, of the $U_4$ spacetime
\cite{Blagojevic2002}.

The action in PGT has the general form 
\begin{equation}
S = \int \di[4]x\; h^{-1}\left[
  \mathcal{L}_{\rm G}\left(\mathcal{R}\idx{^{AB}_{CD}},\mathcal{T}\idx{^A_{BC}}\right)
  + \mathcal{L}_{\rm M}\left(\varphi,\mathcal{D}_A\varphi\right)\right],
\end{equation}
where $h=\mbox{det}(h\idx{_A^\mu})$, $\mathcal{L}_{\rm G}$ is the free
gravitational Lagrangian, $\mathcal{L}_{\rm M}$ is the matter Lagrangian,
$\mathcal{R}\idx{^{AB}_{CD}}\left(h,A,\partial A\right)$ and
$\mathcal{T}\idx{^A_{BC}}(h, \partial h, A)$ are the field strengths
corresponding to the Lorentz and translational parts, respectively, of
the Poincar\'e group, $\mathcal{D}_A$ is the covariant derivative, and
$\varphi$ is the matter field. Here, Greek indices correspond to the
coordinate frame, and capital Latin indices to the local Lorentz frame.
The field strengths can be expressed as
$\mathcal{R}\idx{^{AB}_{CD}}=h\idx{_C^\mu}h\idx{_D^\nu}\mathcal{R}\idx{^{AB}_{\mu\nu}}$
and
$\mathcal{T}\idx{^A_{BC}}=h\idx{_B^\mu}h\idx{_C^\nu}\mathcal{T}\idx{^A_{\mu\nu}}$,
where
\begin{align}
&\mathcal{R}\idx{^{AB}_{\mu\nu}} = 
2(\de_{[\mu}A\idx{^{AB}_{\nu]}}+A\idx{^A_{E[\mu}}A\idx{^{EB}_{\nu]}}),
\label{eqn:Rdef} \\
&\mathcal{T}\idx{^A_{\mu\nu}}=2(\de_{[\mu}b\idx{^A_{\nu]}}
+A\idx{^A_{E[\mu}}b\idx{^E_{\nu]}}),
\end{align}
and $b\idx{^A_\mu}$ is the inverse $h$-field, such that
$b\idx{^A_\mu}h\idx{_B^\mu}=\delta^A_B$ and $b\idx{^A_\mu}h\idx{_A^\nu}=\delta^\nu_\mu$.

One property that a healthy theory should possess is unitarity. The
particle spectrum of a unitary theory should contain no ghosts
(particles with negative free field energy) or tachyons (particles
with imaginary masses). Several authors have previously arrived at
no-ghost-or-tachyon conditions for some subsets of PGT.  For
parity-conserving PGT (which we term PGT$^+$), Neville
\cite{Neville1978,Neville1980} considered $\mathcal{R}+\mathcal{R}^2$
actions, and Sezgin and van Nieuwenhuizen \cite{Sezgin1980} examined
the most general action with no more than two derivatives,
i.e. $\mathcal{R}+\mathcal{R}\text{}\idx{^2}+\mathcal{T}^2$, using a
systematic method with spin projection operators
\cite{Rivers1964,VanNieuwenhuizen1973,Neville1978}. Karananas
\cite{Karananas2015} and Blagojevi\'{c} and Cvetkovi\'{c}
\cite{Blagojevic2018} studied the most general
$\mathcal{R}+\mathcal{R}\text{}\idx{^2}+\mathcal{T}^2$ action for PGT
with parity-violating terms.

If the parameters in the action satisfy certain ``critical conditions'',
however, the theory may possess additional gauge invariances. This
increases the difficulty of obtaining the no-ghost condition of the
massless sector of a PGT systematically. Therefore, following a brief
primer on spin projections operators and notation in
Sec.~\ref{section:primer}, we present in
Sec.~\ref{section:ghost} a systematic approach to investigating all
such critical cases and accommodating the associated additional source
constraints; the method is implemented in \textsc{Mathematica} using
the \textsc{MathGR} \cite{Wang2013PRD} package. We apply our method to
PGT$^+$ in Sec.~\ref{sec:PGTResult} and compare our results with
those previously presented in the literature; we also identify special
cases that are not only free of ghosts and tachyons, but also
power-counting renormalizable.  We conclude in
Sec.~\ref{sec:Conclusion}.

We use the Landau--Lifshitz metric signature $\eta_{AB}=(+,-,-,-)$
throughout this paper.

\section{Spin projection operators \label{section:primer}}

We begin by briefly reviewing the spin projection operator (SPO)
formalism \cite{Rivers1964,Barnes1965,Aurilia1969} and establishing
our notation. The SPOs may be used to decompose a field in momentum
space into parts with definite spin $J$ and parity $P$.

A field $\zeta_{\acute{\alpha}}$, where a Greek index with an
acute accent ($\acute{\alpha}$, ...) represents the collection of the
local Lorentz indices of the field, may be decomposed as
\begin{align}
\zeta(k)_{\acute{\alpha}} &= \sum_{J,P,i} \zeta_i(J^P,k)_{\acute{\alpha}}, \\
\zeta_i(J^P,k)_{\acute{\alpha}} &\equiv
P_{ii}(J^P,k)\idx{_{\acute{\alpha}}^{\acute{\beta}}}
\zeta(k)_{{\acute{\beta}}},
\label{eqn:fielddecomp}
\end{align}
where there is no sum on $i$ in \eqref{eqn:fielddecomp}.
There may be more than one component, or none, with spin-parity
$J^P$. The index $i$ (or, more generally, lowercase Latin letters from
the middle of the alphabet) labels these components in the same
spin-parity sector and also labels the SPOs. The momentum $k^A$ is a
timelike vector, but for simplicity we omit the tensor indices of the
momentum $k$ and position $x$ when they appear as function
arguments. Indeed, for brevity's sake, we will omit the dependence of
fields and SPOs on $k$ or $x$ for the remainder of this section.

There are also off-diagonal SPOs
$P_{ij}(J^P)\idx{_{\acute{\alpha}}^{\acute{\beta}}}$, where $i\neq j$,
which complete a basis for parity-conserving operators acting on
$\zeta_{\acute{\alpha}}$. The SPO basis is Hermitian, complete,
orthonormal, and the diagonal elements are positive (or negative)
definite. Thus, they satisfy
\begin{align}
&P_{ij}(J^P)^{{\acute{\alpha}}{\acute{\beta}}}=P^\ast_{ji}(J^P)^{{\acute{\beta}}{\acute{\alpha}}}, \\
&\sum_{i,J,P}P_{ii}(J^P)_{{\acute{\alpha}}{\acute{\beta}}} = \mathbb{I}_{{\acute{\alpha}}{\acute{\beta}}}, \label{eqn:Pcomplete} \\
&P_{ik}(J^P)\idx{_{\acute{\alpha}}^{\acute{\mu}}}P_{lj}(J^{\prime P'})_{\acute{\mu}{\acute{\beta}}} = \delta_{JJ'}\delta_{PP'}\delta_{kl}P_{ij}(J^P)_{{\acute{\alpha}}{\acute{\beta}}}, \label{eqn:Portho} \\
&[\varphi^{*}_{\acute{\alpha}} P_{ii}(J^P)^{{\acute{\alpha}}{\acute{\beta}}}\varphi_{\acute{\beta}}]P \geq 0 \quad \forall i,\varphi_{\acute{\alpha}}, \label{eqn:PPositiveDefinite}
\end{align}
where $\mathbb{I}_{{\acute{\alpha}}{\acute{\beta}}}$ is the identity
operator for the field $\zeta$, and in the final condition
$\varphi_{\acute{\beta}}$ is an arbitrary field in the same tensor
space as $\zeta$ and $P$ (without indices) is the parity.

Now consider the (usual) case in which the action contains multiple
fields $\zeta^{(1)}_{{\acute{\alpha}}_1}$,
$\zeta^{(2)}_{{\acute{\alpha}}_2}$, \ldots,
$\zeta^{(f)}_{{\acute{\alpha}}_{f}}$, where the index $a=1,\ldots,f$
labels the fields (generally we will use lowercase Latin letter from
the start of the alphabet for this purpose). One can then generalize
the SPO $P_{ij}(J^P)_{{\acute{\alpha}}{\acute{\beta}}}$ in the
single-field case to
$P_{ij}^{(ab)}(J^P)_{{\acute{\alpha}}{\acute{\beta}}}$, where the
latter now projects the $j$th part with spin-parity $J^P$ of the field
$\zeta^{(b)}_{\acute{\beta}}$ into the $i$th part with spin-parity
$J^P$ of the field $\zeta^{(a)}_{\acute{\alpha}}$. 

It is clear from the above discussion that the description of SPOs
requires the introduction of several sets of indices of different
types. In an attempt to ease somewhat this notational burden, we
introduce a matrix-vector formalism that removes two of these sets of
indices. We begin by defining the generalized field vector 
\begin{equation}
\hat{\zeta} \equiv \sum_{a=1}^n \zeta^{(a)}_{\acute{\alpha}_a}
\mathbf{e}_{a},
\label{eqn:fieldvecdef}
\end{equation}
where $\mathbf{e}_{a}$ is a column vector with
$a$th element equal to unity and the remaining elements zero.  On the left-hand side (LHS) of (\ref{eqn:fieldvecdef}), we have suppressed the local Lorentz
indices, and it should be understood that the $a$th element of
$\hat{\zeta}$ consists of the tensor expression
$\zeta^{(a)}_{\acute{\alpha}_a}$.
%
The contraction of two generalized field vectors $\hat{\zeta}$ and $\hat{\xi}$
is then given by
\begin{equation}
\hat{\zeta}^\dagger\cdot \hat{\xi} = \sum_{a=1}^n \zeta^{\ast
  (a)}_{{\acute{\alpha}}_a}\xi^{(a){\acute{\alpha}}_a}, 
\end{equation}
where we have
``overloaded'' the dot notation on the LHS to encompass the summations
both over the field index $a$ and the collection of local Lorentz
indices $\acute{\alpha}$.

Turning to the SPOs, we begin by considering the tensor quantities
$P_{ij}^{(ab)}(J^P)_{{\acute{\alpha}}{\acute{\beta}}}$ as the elements
of a block matrix $\mathsf{P}(J^P)$, for which the indices $(a,b)$
label the $f\times f$ blocks and the indices $(i,j)$ label the
individual elements within each block. Note that since not every field
has parts belonging to a given spin-parity sector $J^P$, some of the
blocks will have zero size. We then redefine the indices $(i,j)$ such
that $P_{ij}(J^P)_{{\acute{\alpha}}{\acute{\beta}}}$ denotes simply
the tensor expression in the $i$th row and $j$th column of
$\mathsf{P}(J^P)$. Finally, for each such element, we define the $f
\times f$ matrix
\begin{equation}
\hat{P}_{ij}(J^P) 
\equiv
P_{ij}(J^P)_{{\acute{\alpha}}{\acute{\beta}}} \mathbf{e}_{a}\mathbf{e}^\dagger_{b},
\label{eqn:spomatdef}
\end{equation}
where $(a,b)$ denotes the block in $\mathsf{P}(J^P)$ to which the
element belongs. By analogy with (\ref{eqn:fieldvecdef}), we have
again suppressed the local Lorentz indices on the LHS of
(\ref{eqn:spomatdef}) for brevity. The advantage of this notation is
that these generalized quantities (denoted by a caret) satisfy
relationships of an analogous form to those given in
\eqref{eqn:Pcomplete}--\eqref{eqn:PPositiveDefinite}.

The SPO block matrices $\mathsf{P}(J^P)$ used in this paper for
PGT$^+$ are listed in Appendix~\ref{sec:SpinProjectionOperator}. One
can obtain the operators for other fields by the method described in
\cite{Aurilia1969}.

\section{Method \label{section:ghost}}

We determine whether a theory contains ghosts or tachyons by adapting
the systematic method of spin projection operators used in
\cite{Sezgin1980,Karananas2015}. We apply the method to
parity-preserving Lagrangians with arbitrary real tensor fields, for
which the linearized Lagrangian can be written as
\begin{align}
\mathcal{L} &= \mathcal{L}_{\rm F}+\mathcal{L}_{\rm I},\nonumber \\
&=\tfrac{1}{2}\sum_{a,b}
\zeta^{(a)}_{{\acute{\alpha}}}(x) 
\mathcal{O}^{(ab)}(\de)^{{\acute{\alpha}}{\acute{\beta}}}
\zeta^{(b)}_{{\acute{\beta}}}(x)\!-\!\sum_{a}\zeta^{(a)}_{\acute{\alpha}}(x)
j^{(a)\acute{\alpha}}(x), \nonumber \\
& = \tfrac{1}{2} \hat{\zeta}^{\rm T}(x) \cdot \hat{\mathcal{O}}(\de)
\cdot \hat{\zeta}(x) -\hat{\zeta}^{\rm T}(x)\cdot \hat{j}(x),
\end{align}
where $\zeta^{(a)}_{\acute{\alpha}}(x)$ are the fields,
$j^{(a)}_{\acute{\alpha}}(x)$ are the corresponding source currents,
and we have defined the generalized operator $\hat{\mathcal{O}}(\de)
\equiv
\mathcal{O}^{(ab)}(\de)^{{\acute{\alpha}}{\acute{\beta}}}\mathbf{e}_a\mathbf{e}_b^\dagger$
(again suppressing local Lorentz indices on the LHS), in which
$\mathcal{O}^{(ab)}(\de)^{{\acute{\alpha}}{\acute{\beta}}}$ is a
polynomial in $\de$ and depends linearly on the coefficients of the
terms in the free-field Lagrangian.

By Fourier transformation, the free-field part of the Lagrangian can
be written
\begin{equation}
\mathcal{L}_{\rm F} = \tfrac{1}{2} \hat{\zeta}^{\rm T}(-k) \cdot
\hat{\mathcal{O}}(k) \cdot \hat{\zeta}(k),
\end{equation}
where, without loss of generality, one may take $\hat{\mathcal{O}}(k)$
to be Hermitian. A theory has no tachyon if
all particles have real masses, and it contains no ghost
particle if the real parts of the residues of the saturated propagator
at all poles are non-negative:
\begin{equation}
\mathrm{Re}\,\left[\underset{k^2=m^2}{\mathrm{Res}}\left( \Pi \right)\right]\geq 0,\label{eqn:noGhostProp}
\end{equation}
where the saturated propagator is the propagator sandwiched between currents
\begin{equation}
\Pi(k) = \hat{j}^\dagger(k) \cdot \hat{\mathcal{O}}^{-1}(k) \cdot \hat{j}(k).
\end{equation}

To obtain the propagator, one first decomposes $\hat{\mathcal{O}}(k)$
into sectors with definite spin and parity:
\begin{equation}
\hat{\mathcal{O}}(k)=\sum_{J,P} \hat{\mathcal{O}}(J^P,k)
= \sum_{i,j,J,P} a_{ij}(J^P,k)\hat{P}_{ij}(J^P,k).
\label{eqn:Odef}
\end{equation}
Pre- and post-multiplying (\ref{eqn:Odef}) by SPOs and using the
orthonormality conditions \eqref{eqn:Portho}, one obtains (omitting
the explicit dependence of quantities on $k$ for brevity)
\begin{align*}
\hat{P}_{ii}(J^P)&\cdot\hat{\mathcal{O}}\cdot \hat{P}_{jj}(J^P) \\
&=\sum_{k,l,J',P'} a_{kl}(J^{\prime P'})\hat{P}_{ii}(J^P)\cdot\hat{P}_{kl}(J'^{P'})\cdot\hat{P}_{jj}(J^P) \\
&=a_{ij}(J^P)\hat{P}_{ij}(J^P), \numberthis
\end{align*}
from which one can read off $a_{ij}(J^P)$ as the coefficient of
$\hat{P}_{ij}(J^P)$. The quantity $a_{ij}(J^P)$ may be considered as
the $(i,j)$th element of a $s \times s$ matrix $a(J^P)$, where
$s$ is the number of parts of spin-parity $J^P$ across all the
fields.

The next step is to invert $\hat{\mathcal{O}}(k)$ to obtain the
propagator. The orthonormality property of the SPO means that
inverting $\hat{\mathcal{O}}(k)$ is equivalent to inverting the
matrices $a(J^P)$. One may, however, find that some of the
$a$-matrices are singular, and so cannot be inverted.

If $a(J^P)$ is singular, then the theory possesses gauge invariances,
as follows. If $a(J^P)$ has dimension $s\times s$ and rank $r$, then
it has $(s-r)$ null right eigenvectors $v_i^{w,R}(J^P)$, where $i$ is
the vector component index and $w$ is a label enumerating the null
eigenvectors (a null eigenvector is an eigenvector that corresponds to
a zero eigenvalue).  Similarly, the transpose matrix $a^{\rm T}(J^P)$
has $(s-r)$ null left eigenvectors $v_i^{w,L}(J^P)$. Thus, if the
generalized field $\hat{\zeta}$ is subjected to a change of the form
\begin{equation}
\delta\hat{\zeta}^w= \sum_{k,J,P}
v_k^{w,R}(J^P)\hat{P}_{kj}(J^P)\cdot 
\hat{\varphi}, \label{eqn:gaugeFreedom}
\end{equation}
where $\hat{\varphi}$ is some arbitrary generalized field, then the
equations of motion $\hat{\mathcal{O}}\cdot \hat{\zeta}=\hat{j}$
remain unchanged.

The null eigenvectors also lead to constraints on the source currents
$\hat{j}$. From the equations of motion, one may show that
\begin{align*}
\sum_{l}v_l^{w,L}(&J^P) \hat{P}_{kl}(J^P)\cdot \hat{j} \\
& = \sum_{l,i,j} v_l^{w,L}(J^P) \hat{P}_{kl}(J^P)\cdot a_{ij}(J^P)\hat{P}_{ij}(J^P)\cdot \hat{\zeta} \\
& = \sum_{i,j}\left[v_i^{w,L}(J^P)a_{ij}(J^P)\right]  \hat{P}_{kj}(J^P)\cdot \hat{\zeta} \\
& = 0 \qquad \forall k, J^P, w. \numberthis \label{eqn:sourceConstraint}
\end{align*}
Hence, one can use the $(s-r)$ field transformations in
\eqref{eqn:gaugeFreedom} to set the corresponding $(s-r)$ parts
$\zeta_{k}(J^P)_{\acute{\alpha}}$ of the field to zero and hence fix
the gauge. This is equivalent to deleting the corresponding $(s-r)$
rows and columns in $a(J^P)$, and thereby $a(J^P)$ becomes
nonsingular (this is most easily implemented by successively
proposing each row/column pair for deletion, and eliminating only
those for which the rank of the matrix is unchanged). We denote the
$a-$matrices after deleting the rows and columns by $b(J^P)$ . Note
that, if the rank of $a(J^P)$ is zero, then there is no particle
content in this spin-parity sector and we will ignore these
spin-parity sectors in the following discussion.

The inverse of $\hat{\mathcal{O}}(J^P)$ then becomes
\begin{equation}
{\hat{\mathcal{O}}^{-1}}(J^P) = \sum_{i,j}b^{-1}_{ij}(J^P)\hat{P}_{ij}(J^P),
\label{eqn:propres}
\end{equation}
where $b^{-1}_{ij}(J^P)$ denotes the $(i,j)$th element of the inverse $b$-matrix,
and the saturated propagator is thus given by
\begin{equation}
\Pi = \sum_{i,j,J,P} b^{-1}_{ij}(J^P)\,\hat{j}^\dagger\cdot\hat{P}_{ij}(J^P)\cdot\hat{j}. \label{eqn:saturatedPropagatorSpinProjector}
\end{equation}

The no-ghost condition \eqref{eqn:noGhostProp} requires us to locate
the poles of the saturated propagator. We first consider those arising
from the elements of the inverse $b$-matrices, which can be written as
\begin{equation}
b^{-1}_{ij}(J^P) = \frac{1}{\mathrm{det}\left[b(J^P)\right]} C^{\rm T}_{ij}(J^P), \label{eqn:inverseB}
\end{equation}
where $C_{ij}(J^P)$ is the cofactor of the element $b_{ij}(J^P)$. Since
$C_{ij}(J^P)$ is polynomial in $k$, all poles of $b^{-1}_{ij}(J^P)$
are located at the zeroes of
$\mathrm{det}\left[b(J^P)\right]$. The determinant in each spin-parity
sector can be written as
\begin{equation}
\mathrm{det}\left[b(J^P)\right] = \alpha k^{2q}(k^2-m_1^2)(k^2-m_2^2)...(k^2-m_r^2), \label{eqn:bDet}
\end{equation}
where $\alpha$ and $m_1,m_2,\ldots,m_q$ (which we assume are nonzero)
are functions of the Lagrangian parameters but independent of $k$,
and $q$ and $r$ are non-negative integers. Thus, $b^{-1}_{ij}(J^P)$ has poles
only at $k^2=0$ and $k^2=m^2_1$, $k^2=m^2_2$,\ldots, $k^2=m^2_r$.

It is worth noting that the reason why there are no odd-order $k$
terms in the determinant is that only the off-diagonal elements of
$b$-matrices contain odd-order $k$ terms. Such an element must belong
to a row and column corresponding to one field with odd indices and
the other with even indices. The odd-order $k$ is always accompanied
by a factor $i$, so such elements are purely imaginary. Since the
$b$-matrix is Hermitian, however, its determinant is real. The terms in odd
powers of $k$ must cancel because they are imaginary, and so the
determinant contains only terms with even powers of $k$.

\subsection{Massless sector}

The no-ghost condition \eqref{eqn:noGhostProp} in the massless sector
is that the residue of the saturated propagator
(\ref{eqn:saturatedPropagatorSpinProjector}) at $k^2=0$ be
non-negative. Besides the poles at $k^2=0$ present in
$b^{-1}_{ij}(J^P)$, the SPOs $P_{ij}(J^P)$ also contain singularities
of the form $k^{-2n}$, where $n$ is a positive integer.

Letting $k^A=(E,\vec{p})$ and $p\equiv\sqrt{\vec{p}^2}$, the particle
energy is given by $E=\sqrt{k^2+p^2}$, and the saturated propagator
can be written (most conveniently in a slightly unorthodox form) as a Laurent series in
$k^2$ in the neighbourhood of $k^2=0$
\begin{equation}
\Pi(k^2,\vec{p})=\sum_{n=-\infty}^N \frac{Q_{2n}} {k^{2n}}, \label{eqn:saturatedPropLaurent}
\end{equation}
where $N$ is an integer and the coefficients $Q_{2n}$ are some
functions of the on-shell momentum $\bar{k}^A\equiv(p,\vec{p})$ and
the on-shell source currents $j^{(a)}_{\acute{\alpha}}(\bar{k})$. If
$N$ is zero or negative, then there is no pole at $k^2=0$ and there is
no propagating massless particle. We will only discuss the $N>0$ cases
here. The no-ghost conditions \eqref{eqn:noGhostProp} are that the
residue of $k^2=0$ be non-negative, so $Q_{2}\geq 0$. Furthermore,
we require that the saturated propagator has a {\em simple} pole in
$k^2$ at this point, since terms proportional to $k^{-2n}$ with $n > 1$
contain ghost states \cite{Heisenberg1957a}. For example, if the
Laurent series of the saturated propagator about $k^2=0$ contains a
term proportional to $k^{-4}$, one can write this as
\begin{equation}
\frac{1}{k^4}=\lim_{\varepsilon\rightarrow 0}\frac{1}{\varepsilon}\left(\frac{1}{k^2}-\frac{1}{k^2+\varepsilon}\right), 
\end{equation}
which contains a normal state and a ghost state. 

To obtain the coefficients $Q_{2n}$ in the Laurent series
(\ref{eqn:saturatedPropLaurent}), one may expand the SPOs
in the saturated propagator, which can then be written 
as a sum of terms of the form
\begin{equation}
\Pi(k)=\sum\frac{\mathcal{P}_1(k^2)\mathcal{C}(k^A,\eta^{AB},j^{(a)}_{{\acute{\alpha}}})}{\mathcal{P}_2(k^2)},
\end{equation}
where $\mathcal{P}_1(k^2)$ and $\mathcal{P}_2(k^2)$ are polynomials of
$k^2$, and $\mathcal{C}(\cdots)$ is a scalar that is obtained from
contracting the tensors in its argument. We require that
$\mathcal{C}(\cdots)$ does not contain the factor $k^2$ because it can
be absorbed into $\mathcal{P}_1(k^2)$. Note that the coefficient
$Q_{2n}$ may not necessarily be given by
\begin{equation}
\sum\left\{\underset{k^2=0}{\mathrm{Res}}\left[ k^{2(n-1)}\frac{\mathcal{P}_1(k^2)}{\mathcal{P}_2(k^2)} \right]\mathcal{C}(k^A,\eta^{AB},j_{\phi,{\acute{\alpha}}})|_{k^2=0}\right\},
\end{equation}
if there exists any nonzero lower-order (smaller $n$) terms because
there may be $k^A$ terms in $\mathcal{C}(\cdots)$. One can accommodate
this situation by expanding the tensor expressions into their
components before taking residues. To this end, it is convenient to
choose a coordinate system such that $k^{A}=(E,0,0,p)$; this greatly simplifies
the calculation without loss of generality, since the saturated
propagator is Lorentz invariant.

Note that the source currents have to satisfy the source constraints
\eqref{eqn:sourceConstraint}. However, \eqref{eqn:sourceConstraint} is
a set of tensor equations, which is difficult to use systematically in
the no-ghost conditions. We thus expand the source constraints into
their components, and then solve the component equation set and
substitute them back to the saturated propagator. Since
\eqref{eqn:sourceConstraint} is a set of homogeneous linear equations,
we can write it in matrix-vector form as
\begin{equation}
{\arraycolsep=1.4pt\def\arraystretch{0.7}
	\mathbf{C}\cdot \mathbf{j}\equiv\left( \begin{array}{cccc}
	c_{11}& c_{12}& \ldots & c_{1q}\\
	\vdots& \ldots& \ldots & \vdots\\
	c_{m1}& c_{m2}& \ldots & c_{mq}\\
	\end{array} \right)\left( \begin{array}{c} j^{(1)}_{0\cdots 0}\\ j^{(1)}_{0\cdots 1}\\ \vdots\\ j^{(f)}_{3\cdots 3}\\\end{array} \right)=0,
}
\end{equation}
where $m$ and $q$ are integers, $c_{ij}$ is the coefficient of the
$j$th component of the source current in the $i$th equation, $f$ is
the total number of fields, and the subscripts of $j^{(i)}$ are
Lorentz indices. The solution is
\begin{equation}
\mathbf{j}=\sum_{i} X_i \mathbf{n}_i, \label{eqn:currentToFreeVar}
\end{equation}
where $\mathbf{n}_i$ are the null vectors of $\mathbf{C}$, and $X_i$
are some free variables.  Note that we have to rescale those null
vectors with factors $(E-p)^n$ in the denominator to avoid introducing
spurious singularities to the saturated propagator, where $n$ is the
minimum integer to make the null vector nonsingular at $E=p$. We then
replace the source current components with $X_i$ using
\eqref{eqn:currentToFreeVar}.

Now the residue only contains the free variables $X_i$, and we can put
them in a column matrix $\mathbf{X}$. The saturated propagator can
then be written as a matrix $\mathbf{M}$ sandwiched by current vectors
$\mathbf{X}$:
\begin{align}
\Pi &= \mathbf{X}^\dagger \cdot \mathbf{M} \cdot \mathbf{X}.  \label{eqn:SatMatrix}
\end{align}
We can also write $Q_{2n}$ in terms of a matrix $\mathbf{Q}_{2n}$ in a
similar way:
\begin{align}
Q_{2n} &= \mathbf{X}^\dagger \cdot \mathbf{Q}_{2n} \cdot \mathbf{X}. \label{eqn:SatMatrixRes}
\end{align}

Since $Q_{2n}=0$ for $n>N$, then $k^{2(N-1)}\Pi$ contains only a
simple pole or no pole at $k^2=0$, and one obtains
\begin{align}
\mathbf{Q}_{2N} =\underset{k^2=0}{\mathrm{Res}}\left[ k^{2(N-1)}\mathbf{M} \right] = \lim_{E\to p}\left[k^{2N} \mathbf{M}\right].
\end{align}
One then calculates the remaining $\mathbf{Q}_{2n}$ by subtracting all
the higher singularities:
\begin{align*}
\mathbf{Q}_{2n}=
 \lim_{E\to p}&\left[k^{2N} \left(\Pi-\sum_{j=n+1}^N\frac{\mathbf{Q}_{2j}}{k^{2j}}\right)\right], \numberthis 
\end{align*}

Thus, we obtain recursively all of the $\mathbf{Q}$-matrices:
$\mathbf{Q}_{2N}$, $\mathbf{Q}_{2(N-1)}$, ..., $\mathbf{Q}_{2}$. For
$\mathbf{Q}_{2n}$ with $n>1$, one requires that each element
in the matrix is zero:
\begin{equation}
\mathbf{Q}_{2n} = 0 \qquad \forall\; p\neq 0, n>1. \label{eqn:masslessHigherCond}
\end{equation}
For $n=1$, corresponding to the $k^{-2}$ pole, the no-ghost condition
is equivalent to requiring that each eigenvalue of $\mathbf{Q}_{2}$ is
non-negative:
\begin{equation}
\textrm{Eigenvalues}(\mathbf{Q}_{2}) \geq 0 \qquad \forall\;p\neq 0. \label{eqn:masslessk2Cond}
\end{equation}
The number of nonzero eigenvalues is equal to the number of degrees
of freedom of the propagating massless particles.

Solving the inequalities in \eqref{eqn:masslessk2Cond} may be quite
time consuming, however, in the cases where the eigenvalues contain some
roots of cubic or even higher polynomials. It is therefore convenient
to convert them into an alternative form. In particular,
if $x_1, \cdots, x_n$ are the roots of a polynomial
$x^n+a_{n-1}x^{n-1}+\cdots+a_{0}=0$ and the roots are guaranteed to be
real, then
\begin{equation}
x_1, \cdots, x_n >0 \Leftrightarrow (-1)^{n-i}a_{i}>0 \qquad \forall\; a_i.
\end{equation}
We can extend the above relation to non-negative roots using the fact
that if there are exactly $z$ zero roots, then
$a_{0},\cdots,a_{z-1}=0$ and $a_z\neq 0$. We then collect the
conditions with 0 to $n$ zero roots. This gives the conditions for
non-negative roots.


\subsection{Massive sector}
In the massive sector, the no-tachyon conditions are simply:
\begin{equation}
m_s^2 > 0 \quad \forall s
\end{equation}
for every spin-parity sector. If this condition is satisfied, one must
then determine if any of the massive particles is a ghost. For
non-tachyonic particles, $k$ is real around $k^2=m_s^2$, and so the
$b$-matrices are Hermitian. Although one can thus expand the saturated
propagator and analyze its poles in a similar manner to that used in
the massless sector, there is a simpler approach in the massive
sector, provided all the masses in all spin sectors are distinct,
which is true in PGT$^+$. We first discuss this case and discuss the
other more general cases later.

From Eqs.~\eqref{eqn:saturatedPropagatorSpinProjector}--\eqref{eqn:bDet}, for
an arbitrary current $\hat{j}$, the no-ghost condition
\eqref{eqn:noGhostProp} may be written as
\begin{widetext}
\begin{align*}
\mathrm{\eqref{eqn:noGhostProp}} \Leftrightarrow 
&\left[\sum_{i,j,J} \frac{1}{\alpha k^{2q}}\left(\prod_{r\neq
    s}\frac{1}{k^2-m_r^2}\right) C^{\rm T}_{ij}(J^P) \hat{j}^\dagger\cdot\hat{P}_{ij}(J^P)\cdot\hat{j}\right]_{k^2=m_s^2}\geq 0 \quad \forall\, \hat{j},s,J,P,\\
\Leftrightarrow &\left[\sum_{i,J} \frac{1}{\alpha
    k^{2q}}\left(\prod_{r\neq s}\frac{1}{k^2-m_r^2}\right)C^{\rm T}_{D,ii}(J^P) \hat{j}_{D}^\dagger\cdot\hat{P}_{ii}(J^P)\cdot\hat{j}_{D}\right]_{k^2=m_s^2}\geq 0 \quad \forall\, \hat{j}_D,s,J,P,\\ \label{eqn:noGhostSpinF1}\numberthis\\
\end{align*}
\end{widetext}
where $C^{\rm T}_{D,ij}(J^P)=\sum_{k,l} U_{ik}(J^P)C^{\rm
  T}_{kl}(J^P)U^{\dagger}_{lj}(J^P)$,
$\hat{j}_{D}=\sum_{i,j}U(J^P)_{ij}\hat{P}(J^P)\idx{_{ij}}\cdot\hat{j}$
and $U(J^P)_{ij}$ are the elements of a unitary matrix of which each
column is a eigenvector of the matrix with elements $C^{\rm T}_{ij}$
(the subscript $D$ thus denotes a diagonal basis). We can write the
last line in \eqref{eqn:noGhostSpinF1} safely because the matrix with
elements $C^{\rm T}_{ij}(J^P,k^2=m_s^2)$ is finite and Hermitian, so
it must have finite real eigenvalues and the transform matrix with
elements $U_{ij}(J^P)$ is finite even at the pole. Since the current
$\hat{j}_D$ is arbitrary and $b^{-1}_{D,ii}(J^P)$ has either no singularity or a simple pole at $k^2=m_s^2$, which we will explain later, then using \eqref{eqn:inverseB} again gives
\begin{align*}
\mathrm{\eqref{eqn:noGhostSpinF1}} \Leftrightarrow &
\sum_{i,J^P} \underset{k^2=m_s^2}{\mathrm{Res}}\left[b^{-1}_{D,ii}(J^P)\right] \\
&\cdot\left[  \hat{j}^\dagger\cdot\hat{P}_{ii}(J^P)\cdot\hat{j}\right]_{k^2=m_s^2}\geq 0 \quad \forall \hat{j},s,J,P. \numberthis \label{eqn:noGhostSpinF2}
\end{align*}
Since $b_{ij}(J^P,k^2)$ is Hermitian for real $k^2$ about $m_s^2>0$, its eigenvalue $b_{D,ii}(J^P,k^2)$ is analytic as
a function of $k^2$ about $m_s^2>0$ \cite{[][{, p. 139}]Kato1982}, and one can Taylor expand
it about $k^2=m_s^2$:
\begin{align*}
b_{D,ii}(J^P,k^2) =& b_{D,ii}(J^P,m_s^2) \\&+ b'_{D,ii}(J^P,m_s^2)\cdot (k^2-m_s^2) + ..., \numberthis\label{eqn:bDiiExpand}
\end{align*}
where the prime denotes the derivative with respect to $k^2$. The
determinant is a polynomial in $k^2$, so it must also be analytic. Since
it equals zero at $k^2=m_s^2$, we can write:
\begin{align*}
\mathrm{det}\left[b(J^P)\right]&(k^2) = \mathrm{det}\left[b(J^P)\right]'(m_s^2)\cdot (k^2-m_s^2) \\&+ \frac{1}{2}\mathrm{det}\left[b(J^P)\right]''(m_s^2)\cdot (k^2-m_s^2)^2 + ... \numberthis
\end{align*}
As we are assuming that all the masses are distinct, then
$\mathrm{det}\left[b(J^P)\right]'(m_s^2)\neq 0$ and
$\mathrm{det}\left[b(J^P)\right](k^2)\sim O(k^2-m_s^2)$ when $k^2$ is
near $m_s^2$. Hence, there should be one $i$ with
$b_{D,ii}(J^P)(m_s^2) \sim O(k^2-m_s^2)$, and the other $b_{D,ii}(J^P)(m_s^2)\sim O(1)$. Thus, exactly one
${\mathrm{Res}}_{k^2=m_s^2}\left[b^{-1}_{D,ii}(J^P)\right]$ is
nonzero. Together with the property \eqref{eqn:PPositiveDefinite},
the massive no-ghost condition therefore becomes
\begin{align*}
\mathrm{\eqref{eqn:noGhostSpinF2}} \Leftrightarrow&  \underset{k^2=m_s^2}{\mathrm{Res}}\left[b^{-1}_{D,ii}(J^P)\right]\cdot P\geq 0 \quad \forall s, \\ 
\Leftrightarrow& \underset{k^2=m_s^2}{\mathrm{Res}}\left[\mathrm{Tr}\,b^{-1}_{D}(J^P)\right]\cdot P\geq 0 \quad \forall s, \\
\Leftrightarrow& \underset{k^2=m_s^2}{\mathrm{Res}}\left[\mathrm{Tr}\,b^{-1}(J^P)\right]\cdot P\geq 0 \quad \forall s. \numberthis \label{eqn:noGhostSpinF3} 
\end{align*}

Let us now examine the case where
${\mathrm{Res}}_{k^2=m_s^2}\left[\mathrm{Tr}\,b^{-1}(J^P)\right]=0$. This
violates the conclusion that exactly one
${\mathrm{Res}}_{k^2=m_s^2}\left[b^{-1}_{D,ii}(J^P)\right]$
is nonzero. The only assumptions we made are that there is no tachyon
and all masses are distinct. Therefore, if
${\mathrm{Res}}_{k^2=m_s^2}\left[\mathrm{Tr}\,b^{-1}(J^P)\right]=0$,
there must be a tachyon or there exist identical masses.

Hence, the combined massive no-ghost-and-tachyon conditions are
\begin{align}
&m_s^2 > 0 \quad \forall s, \label{eqn:noTachyonSpin} \\
&\underset{k^2=m_s^2}{\mathrm{Res}}\left[\mathrm{Tr}\,b^{-1}(J^P)\right]\cdot P> 0 \quad \forall s,  \label{eqn:noGhostSpin}
\end{align}
if the masses in each spin sector are distinct. To obtain the masses,
one merely has to calculate the roots of the determinants of the
$b$-matrices. We assume that all the roots that depend on the
parameters of the Lagrangian are indeed non-zero. If one sets a
nonzero mass to zero, however, a massive pole becomes massless pole
and one has to recalculate the massless no-ghost conditions because
the additional massless pole was not included in the calculation in
the previous previous step. We will discuss such ``critical cases''
later and assume that they do not occur here.

If any mass in a spin sector has multiplicity greater than one,
Eq.~\eqref{eqn:bDiiExpand} will not hold. In that case, one has to
calculate $b^{-1}_{D,ii}(J^P)$ explicitly and use the condition
\eqref{eqn:noGhostSpinF2} directly. One should also avoid higher
singularities in these cases. In the PGT$^+$ case that we consider in
Sec.~\ref{sec:PGTResult}, however, there is at most one massive
mode in each spin sector.

We note that the condition \eqref{eqn:noGhostSpin} is the same as
Eq. (27b) in \cite{Sezgin1980}, but differs from
Eq. (47) in \cite{Karananas2015}. The reason is that
Karananas considers full PGT, with parity-violating terms, so that his
spin projectors do not satisfy
$P_{ij}^*(J^P)_{{\acute{\alpha}}{\acute{\beta}}} =
P_{ji}(J^P)_{{\acute{\beta}}{\acute{\alpha}}}$ and the parity-even and
odd parts are mixed. Hence, \eqref{eqn:noGhostSpinF2} is not valid in
this case. It is not clear, however, how one arrives at Eq. (47)
in \cite{Karananas2015} in the full PGT case.

Finally, we note that the full combination of conditions on the Lagrangian
are given by \eqref{eqn:masslessHigherCond},
\eqref{eqn:masslessk2Cond}, \eqref{eqn:noTachyonSpin} and
\eqref{eqn:noGhostSpin}.

\subsection{Critical cases \label{sec:SpecialCase}}
There are a number of assumptions in the analysis outlined above, so
the process is not complete. To understand this better, let us
reexamine the determinants in \eqref{eqn:bDet}, which can
be written as
\begin{equation}
\mathrm{det}\left[b\left(J^P\right)\right]= k^{2q} \sum_{j=0}^r\left(A_{2j} k^{2j} \right) 
= k^{2q}A_{2r}\prod_{j=1}^r\left(k^2-m_j^2\right), \label{eqn:detAnalyse}
\end{equation}
where $q$ and $r$ are non-negative integers, and $A_{2j}$ are some
finite functions of the parameters, with $A_{2r}\neq0$ and $A_0\neq0$.
In the above process, we have implicitly assumed $m_j\neq 0$ and
finite. We now discuss what may happen if the parameters in the Lagrangian
satisfy some equalities and violate these assumptions in a given
spin-parity sector $J^P$. 

In particular, we consider the following eventualities.
\begin{enumerate}
	\item $\mathrm{det}\left[b\left(J^P\right)\right]=0$: This is
          equivalent to all $A_{2j}=0$. The determinant becomes zero,
          and there are more gauge freedoms. Hence, we need to
          calculate the new source constraints and
          $b^{-1}_{ij}\left(J^P\right)$ matrix elements, and the
          massless, as well as massive poles, have different forms.
	\item \label{itm:degMassiveToMassive}
          $\mathrm{det}\left[b\left(J^P\right)\right]\neq 0$, but
          $A_0=0$: The determinant can then be written as
	\begin{equation}
	\mathrm{det}\left[b\left(J^P\right)\right]= k^{2(q+l)}
        \sum_{j=l}^r\left(A_{2j} k^{2(j-l)} \right),
	\end{equation}
	where $A_{2l} \neq 0$, $l$ is a positive integer and $r\geq
        l>0$. Some masses becomes zero, so some massive poles of the
        propagator become massless. The number of massive conditions
        decreases, and the massless conditions change. Hence, there is
        no further gauge invariance, and the source constraints and
        the matrix elements $b^{-1}_{ij}\left(J^P\right)$ remain in
        the same form. One needs to calculate the new massless and
        massive conditions.
	\item \label{itm:degMassiveRemoved}
          $\mathrm{det}\left[b\left(J^P\right)\right]\neq 0$, and
          $A_{2r}=0$: The second equality of \eqref{eqn:detAnalyse}
          becomes invalid since some masses become infinite. In this
          case, we can write the determinant as
	\begin{equation}
	\mathrm{det}\left[b\left(J^P\right)\right]=k^{2q} \sum_{j=0}^{r-l}\left(A_{2j} k^{2j} \right),
	\end{equation}
	where $l$ is a non-negative integer. There is no new gauge
        freedom, but the number of the roots is decreased. The poles
        are ``removed'' in this case. Since only the $k^{2q}$ part will
        affect the massless poles in the saturated propagator [see
        Eq.~\eqref{eqn:saturatedPropagatorSpinProjector}--\eqref{eqn:bDet}],
        the forms of the massless poles are unchanged. Hence, one need
        only recalculate the massive conditions. In this case, some
        non-propagating modes (propagator with no pole) might
        appear. We do not forbid these modes in this paper.
\end{enumerate}

We can find all conditions that cause a theory to be a critical case
by finding all conditions that cause
$\mathrm{det}\left[b\left(J^P\right)\right]=0$, $A_0=0$, or $A_{2r}=0$
in any spin sectors. While some conditions may cause more than one of
the above situations, we can still divide all the critical conditions
into three categories.
\begin{enumerate}[label=\Alph*.]
	\item Those causing
          $\mathrm{det}\left[b\left(J^P\right)\right]=0$ in any
          spin-parity sector: The source
          constraints, $b^{-1}_{ij}\left(J^P\right)$ matrix elements, and the
          massless as well as massive poles have different forms.
	\item Those causing $A_0=0$ in any spin-parity sector, and not
          belonging to Type A: The form of the source constraints and
          the $b^{-1}_{ij}\left(J^P\right)$ matrix elements are the same, but
          the massless and massive conditions have different forms.
	\item Those conditions not belonging to Type A and Type B:
          These conditions cause $A_{2r}=0$ in some spin sectors. Only
          the form of the massive condition is changed. We can
          substitute the conditions into the massless condition
          directly.
\end{enumerate}

We can then traverse all possible critical cases. First, we find the
type A, B and C conditions for the parameters in the original
Lagrangian satisfying only one equality. Each type A and B condition
is a child theory of the original theory. For the type C conditions, any
combination of type C conditions of a theory is also a type C
condition of the theory, provided they are not contradictory. Note
that we are assuming that a child theory does not satisfy the other
sibling critical conditions, and it does not include the critical
cases of itself. Hence, some combinations of type C conditions might
be contradictory, and we have to remove these cases. We first
calculate the no-ghost-and-tachyon conditions for all the type C child
theories. We then calculate the no-ghost-and-tachyon conditions for
the first type A or B child theory and then find its critical
cases. 

We traverse the ``tree'' in a pre-ordered way: we repeat the above process
until the theory we are investigating has no type A or B child theory,
and then return to its parent theory and consider the next unevaluated child
theory of the parent theory. Because it is possible to reach the same
theory by different routes, we have to check whether the child theory
has been evaluated. If it has been evaluated, we neither calculate it
again nor find its child theories. The reason why we do not have to
find the child theories for type C conditions is that their type A and
B child conditions must be evaluated in some other branches of their
sibling type A or B conditions. As for the type C child theories, they
are already included in the combination of the sibling type C
conditions. We can then find all possible critical cases and collect
all no-ghost-and-tachyon conditions.

This process is best illustrated by examples, which we provide in the
next section, in the context of PGT$^+$.

\section{Application to PGT$^+$ \label{sec:PGTResult}}

The most general free-field PGT$^+$ Lagrangian that is at most 
quadratic in the gravitational gauge fields may be written as:
\begin{align*}
\frac{\mathcal{L}}{b}&=
-\lambda \mathcal{R}
+\left(r_4+r_5\right) \mathcal{R}^{AB} \mathcal{R}_{AB}\\
&+\left(r_4-r_5\right) \mathcal{R}^{AB} \mathcal{R}_{BA} 
+\left(\frac{r_1}{3}+\frac{r_2}{6}\right) \mathcal{R}^{ABCD}
\mathcal{R}_{ABCD}\\
&+\left(\frac{2 r_1}{3}-\frac{2 r_2}{3}\right) 
\mathcal{R}^{ABCD} \mathcal{R}_{ACBD} \\
&+\left(\frac{r_1}{3}+\frac{r_2}{6}-r_3\right) 
\mathcal{R}^{ABCD} \mathcal{R}_{CDAB} \\
&+\left(\frac{\lambda }{4}+\frac{t_1}{3}+\frac{t_2}{12}\right) 
\mathcal{T}^{ABC} \mathcal{T}_{ABC} \\
&+\left(-\frac{\lambda }{2}-\frac{t_1}{3}+\frac{t_2}{6}\right) 
\mathcal{T}^{ABC} \mathcal{T}_{BCA}\\
&+\left(-\lambda -\frac{t_1}{3}+\frac{2 t_3}{3}\right) 
\mathcal{T}\text{}_B\text{}^A\text{}^B \mathcal{T}\text{}_C\text{}_A\text{}^C,
\label{eqn:PGTLagrangian} \numberthis
\end{align*}
where $\mathcal{R}\idx{^A_B}=\mathcal{R}\idx{^{AC}_{BC}}$, $\mathcal{R}=\mathcal{R}\idx{^A_A}$, and we have adopted the conventions in \cite{Sezgin1980} for the
parameters, which simplifies calculations and enables a
straightforward comparison with the literature.

To determine the particle spectrum, one must first linearize the
Lagrangian. We expand it around a Minkowski background with
\begin{equation}
h\text{}_A\text{}^{\mu }=\delta\text{}_A\text{}^{\mu }+f\text{}_A\text{}^{\mu },
\end{equation}
and we set the $A$-field to be $O(f)$. The inverse of $h$ becomes
\begin{equation}
b\text{}^A\text{}_{\mu }=\delta\text{}^A\text{}_{\mu }-f\text{}^{A}\text{}_\mu+O\left(f^2 \right) .
\end{equation}
Since the effect of transforming Greek indices to Latin indices is only $O\left( f^2\right)$, we can ignore the difference between them and only use Latin indices in the linearized theory. We can decompose $f$ into symmetric and antisymmetric parts:
\begin{equation}
f\text{}_A\text{}_B=a\text{}_A\text{}_B+s\text{}_A\text{}_B.
\end{equation}

Note that one may add a constant term $c_0$ to the right-hand side
of \eqref{eqn:PGTLagrangian}, but after the weak field expansion
the Lagrangian becomes
\begin{equation}
\mathcal{L} = c_0 - t\left(2\lambda \de_A A\idx{^B^A_B} + c_0 s\right) + \mathcal{O}\left(t^2\right).
\label{eqn:cosmoconstlin}
\end{equation}
The constant term in \eqref{eqn:cosmoconstlin} does not affect the
equation of motion, so we can neglect it, and the $\de A$ term can be
eliminated by partial integration regardless of whether $c_0$ is
zero. If $c_0\neq 0$, however, the $c_0 s$ term in the
$\mathcal{O}(t)$ part of the Lagrangian results in the equation of
motion $c_0=0$ at order $t$, which contradicts $c_0\neq
0$. Furthermore, we consider only the Minkowski background here, and
adding a cosmological constant term will cause the background to de
Sitter. Hence, $c_0$ must always be zero, and so we do not add the
constant term to  \eqref{eqn:PGTLagrangian}.

Before considering the general case of PGT$^+$, however, we begin by
first studying the simpler cases of PGT$^+$ with vanishing torsion and
curvature, respectively, which one should note are {\em not} merely
critical cases of \eqref{eqn:PGTLagrangian}, because additional
constraints are placed not only the coefficients, but also on the
fields.

\subsection{Zero-torsion PGT$^+$}
\label{sec:ztpgt}

One may impose vanishing torsion as follows \cite{Blagojevic2002}.
First, we define
\begin{align}
c\idx{^A_{\mu\nu}}&\equiv \de_{\mu}b\idx{^A_\nu}-\de_{\nu}b\idx{^A_\mu}\\
\Delta_{AB\mu}&\equiv \frac{1}{2}\left(c_{ABC}-c_{CAB}+c_{BCA}\right)b\idx{^C_\mu},
\end{align}
then the $A$-field can be written as
$A_{AB\mu}=\Delta_{AB\mu}+K_{AB\mu}$,
where $\Delta_{AB\mu}$ are the Ricci rotation coefficients or
``reduced" $A$-field \cite{Lasenby2016}, and
$K_{\mu\lambda\nu}=-\frac{1}{2}\left(\mathcal{T}_{\mu\lambda\nu}-\mathcal{T}_{\nu\mu\lambda}+\mathcal{T}_{\lambda\nu\mu}\right)$
is the contorsion. Hence, setting the torsion to zero is equivalent to
replacing $A_{AB\mu}$ with $\Delta_{AB\mu}$. The Lagrangian of
torsionless PGT is thus
\begin{equation}
\frac{\mathcal{L}}{b}=-\lambda {\mathcal{R}}+2 r_4 
\mathcal{R}^{AB} \mathcal{R}_{AB} +\left(r_1\!-\!r_3\right) 
\mathcal{R}^{ABCD} \mathcal{R}_{ABCD} \label{eqn:TorsionlessPGT}.
\end{equation}

\begin{figure}[h]
	\includegraphics[width=0.48\textwidth]{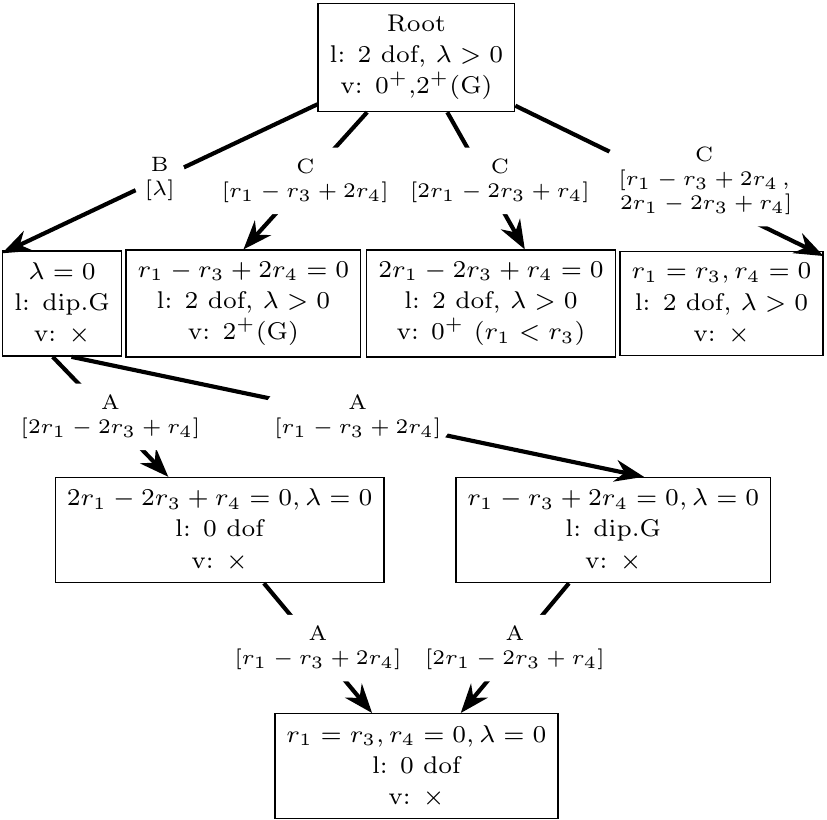}
	\caption{\label{fig:noTorsionTree} The critical cases of
          zero-torsion PGT$^+$, for which the Lagrangian has the form
\eqref{eqn:TorsionlessPGT}. See the text for details.}
\end{figure}

We employ the general method described in Sec.~\ref{section:ghost}
to this case, and present our results Fig.~\ref{fig:noTorsionTree},
which also illustrates our methodology in diagrammatic form. The top
``node'' in the figure (entitled ``root'') represents the full theory
described by \eqref{eqn:TorsionlessPGT}, without imposing any
relationship between the parameters in the Lagrangian. The line ``l''
in each node lists the number of degrees of freedom in the massless
sector and the condition for that sector to be ghost-free;
alternatively it is marked with ``G'' to denote that the sector must
contain a ghost, or ``dip.G'' to denote that it must contain a dipole
ghost. The line ``v'' in each node lists the massive particles and the
conditions that must be satisfied for them to be neither ghosts nor
tachyons; alternatively, it is marked with
a ``G'' if one of them must be a ghost or tachyon. If there is no massive
particle, then $\times$ is written.

The arrows between nodes point from parent theories to their child
theories. The first line of the label on each arrow indicates the type
of the critical case, and the second line denotes that one is setting
the expressions in $\left[...\right]$ to zero in the parent theory to
obtain the child theory. The first line in each node (except the
``root'' node) contains the full set of critical conditions for that
theory.  Note that for each theory, the conditions that make it
critical (the expressions in the arrows from that node) are required
not to hold. For example, for the theory with $\lambda=0$ in the
second row of Fig.~\ref{fig:noTorsionTree}, one requires
$2r_1-2r_3+r_4\neq 0$ and $r_1-r_3+2r_4\neq 0$. The bottom node
corresponds to the Lagrangian vanishing identically. 

For the subset of cases considered previously by other authors, we
compare our results with those in the literature in
Sec.~\ref{sec:comparison}.
%


\subsection{Zero-curvature PGT$^+$}
\label{sec:zcpgt}

\begin{figure}[h]
	\includegraphics[width=0.48\textwidth]{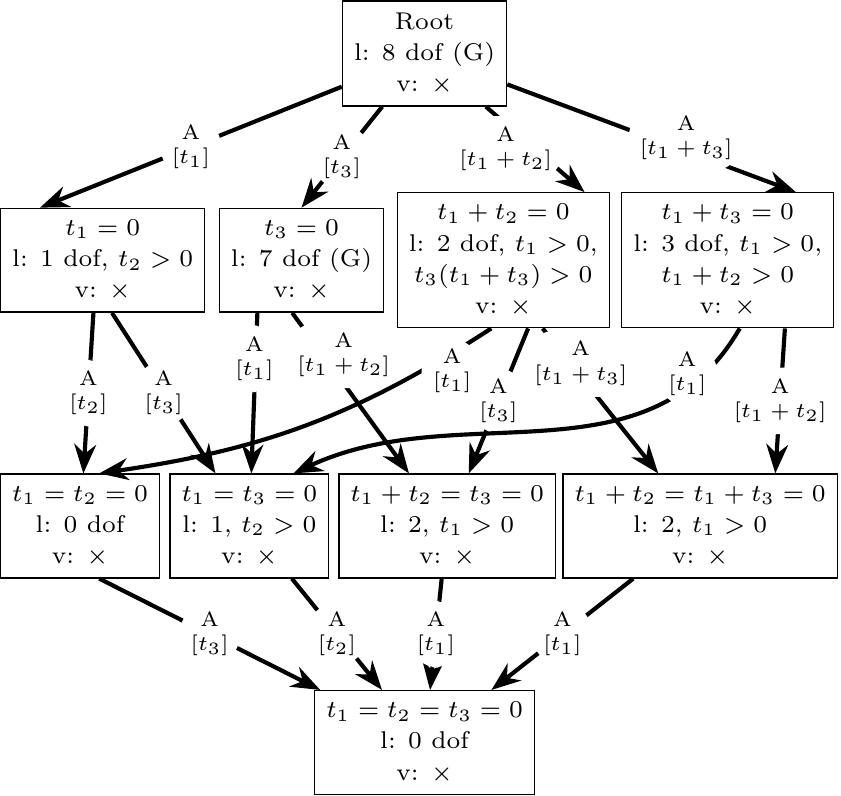}
	\caption{\label{fig:Teleparallel} The critical cases of
          zero-curvature (teleparallel) PGT$^+$, for which the
          Lagrangian has the form \eqref{eqn:TeleparallelPGT}. See
          text for details.}
\end{figure}

One may impose zero curvature in PGT$^+$ (to obtain teleparallel
PGT$^+$) by setting $A_{AB\mu}=0$ \cite{Blagojevic2002}, and the
corresponding Lagrangian is given by
\begin{align*}
\frac{\mathcal{L}}{b}&= \left(\frac{t_1}{3}+\frac{t_2}{12}\right)
\mathcal{T}^{ABC}
\mathcal{T}_{ABC}
+\left(\!-\frac{t_1}{3}+\frac{t_2}{6}\right)
\mathcal{T}^{ABC}
\mathcal{T}_{BCA}\\
&+\left(\!-\frac{t_1}{3}+\frac{2t_3}{3}\right) 
\mathcal{T}\text{}_B\text{}_A\text{}^B
\mathcal{T}\text{}_C\text{}^A\text{}^C.
\label{eqn:TeleparallelPGT}
\numberthis
\end{align*}
Applying the method described in Sec.~\ref{section:ghost} to this
case yields the results presented in results
Fig.~\ref{fig:Teleparallel}, which uses the same conventions as in
Fig.~\ref{fig:noTorsionTree}. We again compare our results with the
literature in Sec.~\ref{sec:comparison}.

\subsection{Full PGT$^+$}

We now turn our attention back to the general case of full PGT$^+$,
for which the Lagrangian is given by
\eqref{eqn:PGTLagrangian}. Starting from the ``root'' theory, for which
no relationship is imposed on the parameters in the Lagrangian, our
method outlined in Sec.~\ref{section:ghost} systematically
identifies 1918 critical cases (excluding the ``vanishing'' Lagrangian
case for which all parameters are zero), which thus cannot be
displayed in diagrammatic form such as in Figs~\ref{fig:noTorsionTree}
and \ref{fig:Teleparallel}. Of these critical cases, we find
that 450 can be free of ghosts and tachyons, provided the parameters
in each case satisfy some conditions without generating another
critical case. The full set of results displayed in an interactive
form can be found at:
\url{http://www.mrao.cam.ac.uk/projects/gtg/pgt/}.

\subsection{Comparison with previous results}
\label{sec:comparison}

We content ourselves here with presenting in
Table~\ref{tab:PGTCompareResult} our results for the root PGT$^+$
theory and the small subset of critical cases that have been studied
previously in the literature. We also list those critical cases of the
torsionless and teleparallel PGT$^+$ theories (see
Figs.~\ref{fig:noTorsionTree} and \ref{fig:Teleparallel}) that have
been considered previously in the literature. Overall, we find that
our results are indeed consistent with those reported by other
authors, apart from a few minor differences that are most likely the
result of typographical errors in earlier papers.

\def\arraystretch{1.9}
\begin{table*}[t!]
	\caption{\label{tab:PGTCompareResult} Conditions for no ghosts
          or tachyons for the PGT$^+$ root theory and a subset
          of critical cases analyzed previously in the literature.
``Massless/massive'' denotes the particle content found in
          the literature, and the parentheses contain the number of 
          degrees of freedom of particles in the massless
          sector. ``Dip. G'' means the massless sector contains
          a dipole ghost. Where our results differ from those in the literature,
          ours are put in squared brackets. Cells marked with
          ``*'' are discussed further in the text, and ``-'' means the
          particle content is not mentioned in the cited paper.}
\begin{tabular}{l@{\extracolsep{0.3cm}}cllcc}
	\toprule
	\#&Paper&Critical Conditions&No-ghost-and-tachyon Conditions&Massless&Massive\\
	\colrule\vspace{3pt}
\refrownumber{row:Root} & \cite{Sezgin1980}& $\times$ & Ghost (massive) & $2^+$ (2) & \makecell[cl]{$0^-$,$0^+$,$1^-$,\\ $1^+$,$2^-$,$2^+$}\\
		\rownumber & \cite{Sezgin1980}& $t_1=t_2=t_3=r_1=r_2=r_3=r_4=r_5=0$ & $\lambda > 0$ & $2^+$ (2) & $\times$ \\
		\refrownumber{row:ECSezgin1980} & \cite{Sezgin1980}& \makecell[cl]{$t_1=-t_2=-t_3=-\lambda$,\\ $r_1=r_2=r_3=r_4=r_5=0$} & $\lambda > 0$ & $2^+$ (2) & $\times$\\
		\rownumber & \cite{Sezgin1980}& \makecell[cl]{$t_1=-t_2=-t_3=-\lambda$, \\ $r_1=r_3=r_4=r_5=0$} & $\lambda > 0, r_2<0$ & $2^+$ (2) & $0^-$\\
		\refrownumber{row:SignDiffSezgin1980} & \cite{Sezgin1980}& \makecell[cl]{$t_1=-t_2, r_1=r_3$, $r_4=r_2=0$} & \makecell[cl]{$\lambda>0, r_1>0 [<0]$, $r_1+r_5>0 [<0]$,\\ $t_1>0, t_3(t_1+t_3)>0$} & $2^+$ (2) & $1^-$, $2^-$\\
		\refrownumber{row:TorsionFreeSezgin1980}& \cite{Sezgin1980}& $t_1=-t_2, r_1=r_3$, $r_4=r_2=0$, torsionless & $\lambda>0$ & $2^+$ (2) & $\times$ \\
		\rownumber & \cite{Sezgin1980}& \makecell[cl]{$t_1=-t_3=-\lambda, r_1=0$, $r_4=-r_5$} & \makecell[cl]{$t_2>\lambda>0, r_2<0$, $2r_3+r_5>0$} & $2^+$ (2) & $0^-$, $1^+$\\
		\rownumber & \cite{Sezgin1980}& \makecell[cl]{$t_1=-t_3=-\lambda, r_1=0$,\\ $r_4=-r_5, r_2=0$} & \makecell[cl]{$2r_3+r_5>0, \lambda>0$, $t_2(t_2- \lambda)>0$} & $2^+$ (2) & $1^+$\\
		\rownumber & \cite{Sezgin1980}& \makecell[cl]{$t_1=-t_3=-\lambda, r_1=0$,\\ $r_4=-r_5, r_2=0$, torsionless} & Ghost & - & -\\
		\rownumber & \cite{Sezgin1980}& $r_1=0, 2r_3=r_4=-r_5$ & \makecell[cl]{$\lambda>0, r_2<0, r_3>0,t_2>0$,\\$t_3 \left(\lambda -t_3\right)<0$} & $2^+$ (2) & $0^-$, $0^+$\\
		\rownumber & \cite{Sezgin1980,Stelle1978}&
                \makecell[cl]{$r_1=0, 2r_3=r_4=-r_5$, torsionless} & $\lambda>0,r_3>0$ & - [(2)] & - $[0^+]$\\
		\refrownumber{row:Sezgin1981TypeC} & \cite{Sezgin1981}& (1)-(12)* & * & $2^+$ (2) & *\\
		\refrownumber{row:NewGISezgin1981} & \cite{Sezgin1981}& \makecell[cl]{$t_1=t_2=t_3=0, r_1=r_3$, \\$r_4=0, 2r_3+r_5=0$} & \makecell[cl]{$\lambda>0,r_1>0$\\$[\lambda>0]$*} & \makecell[cc]{$2^+$,$1^-$*\\ (4) [(2)]} & $\times$\\
		\refrownumber{row:Kuhfuss1986} & \cite{Kuhfuss1986}& \makecell[cl]{$t_1=t_2=t_3=0, r_1=r_3$, \\$r_4=0, 2r_3+r_5=0$} & $\lambda>0$* & $2^+$ (2)& $\times$\\
		\refrownumber{row:teleparallelKuhfuss1986} & \cite{Kuhfuss1986}& \makecell[cl]{$t_1=-t_3$, teleparallel} & $t_1+t_2>0,t_1+\lambda>0$* & $2^+$,$0^+$ (3)& $\times$\\
		\refrownumber{row:teleparallelLord1988} & \cite{Lord1988}& \makecell[cl]{$t_1=-t_3$, teleparallel} & $t_1+\lambda>0$ & \makecell[cc]{$2^+$ (2) \\ $[(3)]$}& $\times$\\
		\rownumber & \cite{Battiti1985}& $r_4=-(r_1/2)+r_3/2, t_3=0$ & \makecell[cl]{$r_1+r_3+2r_5<0$, $\lambda>0$ (massless)} & $2^+$,$1$ (4)& -\\
		\rownumber & \cite{Battiti1985}& $r_2=0, t_2=0$ & \makecell[cl]{$2 r_3+r_5>0, \lambda>0$ (massless)} & $2^+$,$1$ (4)& -\\
		\refrownumber{row:Battiti1985No3} & \cite{Battiti1985}& $t_2=t_3=r_1-r_3+2r_4=r_2=0$ & \makecell[cl]{$2r_3+r_5>0$, $r_1+r_3+2r_5<0$,\\ $\lambda>0$ (massless)} & $2^+$,$1$,$1$ (6)& -\\
		\refrownumber{row:TorsionlessRoot} & \cite{Stelle1978}& torsionless & \makecell[cl]{Ghost (massive $2^+$)} & $2^+$ (2)& $0^+$,$2^+$\\
		\refrownumber{row:BiswasConformal} & \cite{Biswas2013,Stelle1978}& $r_1-r_3+2r_4=0$, torsionless & \makecell[cl]{Ghost (massive $2^+$)} & $2^+$ (2)& $2^+$\\
		\refrownumber{row:NormalConformal} &
                \cite{Riegert1984}& $r_1-r_3+2r_4=\lambda=0$,
                torsionless & \makecell[cl]{Ghost (massless)} &
                \makecell[cl]{$2^+$,$1$,$2^+$\\ (dip. G)*}& $\times$ \\
\botrule
\end{tabular}
\end{table*}

\setcounter{magicrownumbers}{0} 
Some of the cases listed in Table~\ref{tab:PGTCompareResult} are worthy of further discussion, as follows:
\begin{itemize}
	\item Case~\ref{row:Root}: This is the ``root'' PGT$^+$ theory,
          in which no critical condition holds. We find the massless
          no-ghost condition $\lambda > 0 $, which agrees with
          \cite{Sezgin1980}. In the massive case, we find
          the no-tachyon condition in each spin-parity sector to be:
	\begin{align*}
	0^-&: -\frac{t_2}{r_2} > 0
	\\
	0^+&: \frac{t_3 \lambda }{2 \left(r_1-r_3+2 r_4\right) \left(t_3-\lambda \right)} > 0
	\\
	1^-&: -\frac{3 t_1 t_3}{2 \left(r_1+r_4+r_5\right) \left(t_1+t_3\right)} > 0
	\\
	1^+&: -\frac{3 t_1 t_2}{2 \left(2 r_3+r_5\right) \left(t_1+t_2\right)} > 0
	\\
	2^-&: -\frac{t_1}{2 r_1} > 0
	\\
	2^+&: -\frac{t_1 \lambda }{2 \left(2 r_1-2 r_3+r_4\right) \left(t_1+\lambda \right)} >0, \numberthis
	\end{align*}
	and the no-ghost condition in each sector is:
	\begin{align*}
	0^-&:-\frac{1}{r_2}>0
	\\
	0^+&:\frac{-r_1 t_3+r_3 t_3-2 r_4 t_3-t_3 \lambda +\lambda ^2}{2 \left(r_1-r_3+2 r_4\right) \lambda  \left(-t_3+\lambda \right)}>0
	\\
	1^-&:-\frac{3 \left(t_1^2+2 t_3^2\right)}{2 \left(r_1+r_4+r_5\right) \left(t_1+t_3\right){}^2}>0
	\\
	1^+&:\frac{3 \left(t_1^2+2 t_2^2\right)}{2 \left(2 r_3+r_5\right) \left(t_1+t_2\right){}^2}>0
	\\
	2^-&:-\frac{1}{r_1}>0
	\\
	2^+&:\frac{-2 r_1 t_1+2 r_3 t_1-r_4 t_1+t_1 \lambda +\lambda
          ^2}{\left(2 r_1-2 r_3+r_4\right) \lambda  \left(t_1+\lambda
          \right)}>0. \numberthis
	\end{align*}
	These conditions are again equivalent to those in
        \cite{Sezgin1980}, as expected, and cannot be
        satisfied simultaneously. Hence, the theory contains a massive
        ghost, as is well known.
	\item Case~\ref{row:ECSezgin1980}: This is Einstein--Cartan
          theory, and our results are consistent with the literature.
	\item Case~\ref{row:SignDiffSezgin1980}: Our conditions
          $\lambda>0, r_1<0, r_1+r_5<0, t_1>0, t_3(t_1+t_3)>0$ differ
          from the conditions $\lambda>0, r_1>0, r_1+r_5>0, t_1>0,
          t_3(t_1+t_3)>0$ found in \cite{Sezgin1980} in that
          two of the inequalities have the opposite sign. We believe
          these are typos in \cite{Sezgin1980}.
	\item Case~\ref{row:TorsionFreeSezgin1980}: This torsionless
          theory corresponds to that in node 4 of row 2 in
          Fig.~\ref{fig:noTorsionTree}. We obtain the condition
          $\lambda>0$, with only 2 massless degrees of freedom, but
          \cite{Sezgin1980} also set $2t_3-t_1=3\lambda,
          r_5=0$. These additional conditions neither cause the theory
          to become a critical case nor contradict the other
          conditions, so adding them has no effect on the particle
          content.  \cite{Sezgin1980} finds that the action
          reduces to the Einstein action, which is consistent with our
          result.
	\item Case~\ref{row:Sezgin1981TypeC}: We find that the
          critical cases that contain three coefficient equations and
          only type C critical conditions are precisely the 12
          cases listed in Table I of \cite{Sezgin1981}, and
          we obtained the same particle content for each theory.
	\item Case~\ref{row:NewGISezgin1981}: Our no-ghost conditions
          and massless particle content are different from those found
          in \cite{Sezgin1981}. However,
          \cite{Kuhfuss1986} studied the same theory and
          obtained the same conditions and particle content as
          ours. Moreover, our result that there is no massless
          propagating tordion in this theory is also found in 
          \cite{Blagojevic1987}. We notice that, compared
          to  our analysis, some
          terms in Eq. (8) in \cite{Sezgin1981} have
          different signs, which we believe to be typos.
	\item Case~\ref{row:Kuhfuss1986} and
          \ref{row:teleparallelKuhfuss1986}: We find that there is an
          overall sign difference between our linearized Lagrangian
          and that in \cite{Kuhfuss1986}, so the conditions
          also have an overall sign difference. We assume that this is
          a minor error either in their calculation or our conversion
          of it to our notation. We have thus added an overall minus
          sign to their conditions.
	\item Case~\ref{row:teleparallelKuhfuss1986}: This theory was
          also studied in \cite{Lord1988} (along with
          Case~\ref{row:teleparallelLord1988}), who found only a
          spin-2 massless mode with the condition
          $t_1+\lambda>0$. However, they studied only the spin-2
          particles, so our results are consistent.
	\item Case~\ref{row:Battiti1985No3}: We believe that the
          condition $\alpha-\gamma_3=0$ quoted in
          \cite{Battiti1985} contains a typo and should
          instead read $\alpha-\gamma_3\neq0$, which
          is equivalent to $t_1=0 \to t_1\neq 0$ in our notation, thus
          yielding our result.
	\item Case~\ref{row:NormalConformal}: This is conformal
          gravity. \cite{Riegert1984} showed it has a
          normal spin-2, a normal spin-1, and a ghost spin-2 mode, all
          massless. We find there is no massive mode, and there must
          be dipole ghost(s) in the massless sector. Our method can
          determine the existence of ghosts, but not the degrees of
          freedom in the massless sector if there are dipole
          ghost(s). Nonetheless, the results are consistent.
\end{itemize}

\subsection{Source constraints}

As mentioned previously, if the parameters in the PGT$^+$ Lagrangian
\eqref{eqn:PGTLagrangian} satisfy some specific conditions (type A
critical cases), then the resulting theory may possess extra gauge
invariances beyond the Poincar\'e symmetry assumed in its
construction.  For example, for Case~\ref{row:NewGISezgin1981} in
Table~\ref{tab:PGTCompareResult}, it is noted in \cite{Sezgin1981,Kuhfuss1986} that
the theory is additionally invariant under the gauge transformation
\begin{align}
\delta A_{ABC} &= \de_{A}\Lambda_{BC} - \de_{B}\Lambda_{AC} + \de_{C}
\theta_{AB},\label{eqn:SezginGI}
\end{align}
where $\de^{B}\Lambda_{AB}=0$,
$\theta_{AB}=\de_{A}V_{B}-\de_{B}V_{A}$, $\de^{A}V_{A}=0$ and
$\Lambda$ and $V$ are arbitrary (see also \cite{Blagojevic1987}), and has the additional source constraints
\begin{align}
\de^{B} \tau_{ABC} &= 0,\qquad \de^{C} \tau_{ABC} = 0,
\end{align}
beyond the standard ones $\partial^B\sigma_{AB} = 0$ and
$\sigma_{[AB]} + \partial^C\tau_{ABC} = 0$ arising from the Poincar\'e
symmetry. Here, $\sigma_{AB}$ and $\tau_{ABC}$ are the source currents
of the $f_{AB}$ (graviton) and $A_{ABC}$ (tordion) gravitational fields,
respectively.

Our approach also found the same source constraints for this theory,
although not directly as tensor equations, but instead in component
form for $k$ aligned with the $z$-direction. Indeed, we found there are 310
different sets of source constraints among the root PGT$^+$ theory and
its 1918 critical cases.  We are not able to convert all of them
automatically into their corresponding tensor equations, but it is
possible to make such a conversion in some cases. This is performed by
first suggesting possible tensor equations from the patterns present
in the component equations, then converting the possible tensor
equations into component forms, and finally comparing whether they are
equivalent. In \cref{tab:someSC}, we present the results for all
the sets of sources constraints that we were able to convert into
tensor form. We find that the same set of source constraints may hold
for more than one critical case, so in the table we list only the case
having the simplest critical conditions.  It is worth noting that the
first case listed is the root PGT$^+$ theory, for which we recover the
two well-known source constraints arising from the Poincar\'e symmetry
alone. We also note that, aside from the root theory, the numbering of
cases in the table is not related to that used in
\cref{tab:PGTCompareResult}.

\bgroup
\def\arraystretch{1.3}
\begin{longtable*}{@{\extracolsep{0.2cm}}>{\centering\arraybackslash}p{0.025\textwidth}p{0.35\textwidth}@{\extracolsep{0.6cm}}p{0.53\textwidth}}
	\caption{\label{tab:someSC} Source constraints for the root
          PGT$^+$ theory and those critical cases for which the
          constraints could be found in tensor form. Note that 
there may be more than one critical case sharing the same source
constraints, so we list 
only the case having the simplest critical conditions. The numbering
of cases is not related to that used in \Cref{tab:PGTCompareResult}.}\\*
	\noalign{\vspace{3pt}}%
	\toprule
	\# & Critical Conditions & Source constraints\\*
	\colrule
	\noalign{\vspace{3pt}}%
	\endfirsthead
	\multicolumn{3}{c}{TABLE~\ref{tab:someSC} (continued): Source constraints for some PGT$^+$ critical cases.}
	\rule{0pt}{12pt}\\
	\noalign{\vspace{1.5pt}}
	\colrule\rule{0pt}{12pt}
	\# & Critical Conditions & Source constraints\\*
	\colrule
	\noalign{\vspace{3pt}}%
	\endhead
	\noalign{\nobreak\vspace{3pt}}%
	\colrule
	\endfoot
	\noalign{\nobreak\vspace{3pt}}%
	\botrule
	\endlastfoot
	
\rownumber & $\times$ & $k\text{}^{B}\sigma \text{}_{AB}=\sigma \text{}_{AB}-\sigma \text{}_{BA}+2 i k\text{}^{C}\tau \text{}_{ABC}=0$ \\
\rownumber & $r_1-r_3=r_4=\lambda =0$ & $i \sigma \text{}_{AB}+i \sigma \text{}_{BA}-2 k\text{}^{C}\tau \text{}_{CAB}-2 k\text{}^{C}\tau \text{}_{CBA}=i \sigma \text{}_{AB}-i \sigma \text{}_{BA}-2 k\text{}^{C}\tau \text{}_{ABC}=0$ \\
\rownumber & $r_1/2-r_3/2+r_4=r_1/2+r_3/2+r_5=t_3=0$ & $k\text{}^{B}\sigma \text{}_{AB}=\sigma \text{}_{AB}-\sigma \text{}_{BA}+2 i k\text{}^{C}\tau \text{}_{ABC}=g\text{}^{BC}\tau \text{}_{ACB}=0$ \\
\rownumber & $r_1/2-r_3/2+r_4=r_1/2+r_3/2+r_5=t_1=t_3=0$ & $k\text{}^{B}\sigma \text{}_{AB}=\sigma \text{}_{AB}-\sigma \text{}_{BA}+2 i k\text{}^{C}\tau \text{}_{ABC}=k\text{}^{B}\sigma \text{}_{BA}=g\text{}^{BC}\tau \text{}_{ACB}=0$ \\
\rownumber & $r_1=r_3=r_4=r_5=t_1=t_3=0$ & $k\text{}^{B}\sigma \text{}_{AB}=\sigma \text{}_{AB}-\sigma \text{}_{BA}-2 i k\text{}^{C}\tau \text{}_{CBA}=\tau \text{}_{ACB}+\tau \text{}_{BCA}=0$ \\
\rownumber & $r_1=r_3=r_4=r_5=t_1=t_2=t_3=0$ & $k\text{}^{B}\sigma \text{}_{AB}=\sigma \text{}_{AB}-\sigma \text{}_{BA}=k\text{}^{C}\tau \text{}_{CBA}=\tau \text{}_{ACB}+\tau \text{}_{BCA}=0$ \\
\rownumber & $r_1=r_3=r_4=r_5=t_1=t_2=t_3=\lambda =0$ & $\sigma \text{}_{AB}=k\text{}^{C}\tau \text{}_{CBA}=\tau \text{}_{ACB}+\tau \text{}_{BCA}=0$ \\
\rownumber & $r_1=r_2=r_3=r_4=r_5=t_1=t_2=t_3=0$ & $k\text{}^{B}\sigma \text{}_{AB}=\sigma \text{}_{AB}-\sigma \text{}_{BA}=\tau \text{}_{BCA}=0$ \\
\rownumber & $r_1=r_3=r_4=r_5=t_1=t_3=\lambda =0$ & $\sigma \text{}_{AB}-i k\text{}^{C}\tau \text{}_{CBA}=\tau \text{}_{ACB}+\tau \text{}_{BCA}=0$ \\
\rownumber & $r_1/3-r_3=r_1/3+r_4=2 r_1/3+r_5=t_1=t_3=0$ & $k\text{}^{B}\sigma \text{}_{AB}=\sigma \text{}_{AB}-\sigma \text{}_{BA}+i k\text{}^{C}\tau \text{}_{CAB}-i k\text{}^{C}\tau \text{}_{CBA}=g\text{}^{BC}\tau \text{}_{ACB}=2 k\text{}^{C}\tau \text{}_{ABC}-k\text{}^{C}\tau \text{}_{CAB}+k\text{}^{C}\tau \text{}_{CBA}=0$ \\
\rownumber & $r_1/3-r_3=r_1/3+r_4=2 r_1/3+r_5=t_1=t_2=t_3=0$ & $k\text{}^{B}\sigma \text{}_{AB}=\sigma \text{}_{AB}-\sigma \text{}_{BA}=g\text{}^{BC}\tau \text{}_{ACB}=k\text{}^{C}\tau \text{}_{CAB}-k\text{}^{C}\tau \text{}_{CBA}=k\text{}^{C}\tau \text{}_{ABC}=0$ \\
\rownumber & $r_2=r_1/3-r_3=r_1/3+r_4=2 r_1/3+r_5=t_1=t_2=t_3=0$ & $k\text{}^{B}\sigma \text{}_{AB}=\sigma \text{}_{AB}-\sigma \text{}_{BA}=g\text{}^{BC}\tau \text{}_{ACB}=k\text{}^{C}\tau \text{}_{CAB}-k\text{}^{C}\tau \text{}_{CBA}=\tau \text{}_{ABC}-\tau \text{}_{ACB}+\tau \text{}_{BCA}=0$ \\
\rownumber & $r_2=r_1/3-r_3=r_1/3+r_4=2 r_1/3+r_5=t_1=t_2=t_3=\lambda =0$ & $\sigma \text{}_{AB}=g\text{}^{BC}\tau \text{}_{ACB}=k\text{}^{C}\tau \text{}_{CAB}-k\text{}^{C}\tau \text{}_{CBA}=\tau \text{}_{ABC}-\tau \text{}_{ACB}+\tau \text{}_{BCA}=0$ \\
\rownumber & $r_1/3-r_3=r_1/3+r_4=2 r_1/3+r_5=t_1=t_2=t_3=\lambda =0$ & $\sigma \text{}_{AB}=g\text{}^{BC}\tau \text{}_{ACB}=k\text{}^{C}\tau \text{}_{CAB}-k\text{}^{C}\tau \text{}_{CBA}=k\text{}^{C}\tau \text{}_{ABC}=0$ \\
\rownumber & $r_1/3-r_3=r_1/3+r_4=2 r_1/3+r_5=t_1=t_3=\lambda =0$ & $\sigma \text{}_{AB}+i k\text{}^{C}\tau \text{}_{ABC}=g\text{}^{BC}\tau \text{}_{ACB}=2 i \sigma \text{}_{AB}-k\text{}^{C}\tau \text{}_{CAB}+k\text{}^{C}\tau \text{}_{CBA}=0$ \\
\rownumber & $r_1-r_3=r_4=r_1+r_5=t_1=t_2=t_3=\lambda =0$ & $\sigma \text{}_{AB}=g\text{}^{BC}\tau \text{}_{ACB}=k\text{}^{C}\tau \text{}_{CAB}+k\text{}^{C}\tau \text{}_{CBA}=k\text{}^{C}\tau \text{}_{ABC}=0$ \\
\rownumber & $r_1-r_3=r_4=r_1+r_5=t_1=t_3=0$ & $k\text{}^{B}\sigma \text{}_{AB}=\sigma \text{}_{AB}-\sigma \text{}_{BA}+2 i k\text{}^{C}\tau \text{}_{ABC}=g\text{}^{BC}\tau \text{}_{ACB}=k\text{}^{C}\tau \text{}_{CAB}+k\text{}^{C}\tau \text{}_{CBA}=0$ \\
\rownumber & $r_1-r_3=r_4=r_1+r_5=t_1=t_3=\lambda =0$ & $\sigma \text{}_{AB}+i k\text{}^{C}\tau \text{}_{ABC}=g\text{}^{BC}\tau \text{}_{ACB}=k\text{}^{C}\tau \text{}_{CAB}+k\text{}^{C}\tau \text{}_{CBA}=0$ \\
\rownumber & $r_1-r_3=r_4=r_1+r_5=t_1=t_2=t_3=0$ & $k\text{}^{B}\sigma \text{}_{AB}=\sigma \text{}_{AB}-\sigma \text{}_{BA}=g\text{}^{BC}\tau \text{}_{ACB}=k\text{}^{C}\tau \text{}_{CAB}+k\text{}^{C}\tau \text{}_{CBA}=k\text{}^{C}\tau \text{}_{ABC}=0$ \\
\rownumber & $r_1/2-r_3/2+r_4=r_1/2+r_3/2+r_5=t_1=t_2=t_3=0$ & $k\text{}^{B}\sigma \text{}_{AB}=\sigma \text{}_{AB}-\sigma \text{}_{BA}=g\text{}^{BC}\tau \text{}_{ACB}=k\text{}^{C}\tau \text{}_{ABC}=0$ \\
\rownumber & $r_2=r_1-r_3=r_4=2 r_1+r_5=t_1=t_2=t_3=\lambda =0$ & $\sigma \text{}_{AB}=k\text{}^{C}\tau \text{}_{CBA}=\tau \text{}_{ABC}-\tau \text{}_{ACB}+\tau \text{}_{BCA}=0$ \\
\rownumber & $r_1/2-r_3/2+r_4=r_1/2+r_3/2+r_5=t_1=t_2=t_3=\lambda =0$ & $\sigma \text{}_{AB}=g\text{}^{BC}\tau \text{}_{ACB}=k\text{}^{C}\tau \text{}_{ABC}=0$ \\
\rownumber & $r_1-r_3=r_4=2 r_1+r_5=t_1=t_2=t_3=\lambda =0$ & $\sigma \text{}_{AB}=k\text{}^{C}\tau \text{}_{CBA}=k\text{}^{C}\tau \text{}_{ABC}=0$ \\
\rownumber & $r_1/2-r_3/2+r_4=r_1/2+r_3/2+r_5=t_1=t_3=\lambda =0$ & $k\text{}^{B}\sigma \text{}_{AB}=g\text{}^{BC}\tau \text{}_{ACB}=i \sigma \text{}_{AB}-k\text{}^{C}\tau \text{}_{ABC}=0$ \\
\rownumber & $r_1-r_3=r_4=2 r_1+r_5=t_1=t_3=\lambda =0$ & $\sigma \text{}_{AB}-i k\text{}^{C}\tau \text{}_{CBA}=2 \sigma \text{}_{AB}+i k\text{}^{C}\tau \text{}_{ABC}+i k\text{}^{C}\tau \text{}_{CAB}=0$ \\
\rownumber & $r_1-r_3=r_4=r_1+r_5=t_3=\lambda =0$ & $g\text{}^{BC}\tau \text{}_{ACB}=i \sigma \text{}_{AB}+i \sigma \text{}_{BA}-2 k\text{}^{C}\tau \text{}_{CAB}-2 k\text{}^{C}\tau \text{}_{CBA}=i \sigma \text{}_{AB}-i \sigma \text{}_{BA}-2 k\text{}^{C}\tau \text{}_{ABC}=0$ \\
\rownumber & $r_1-r_3=r_4=t_1=t_2=t_3=\lambda =0$ & $\sigma \text{}_{AB}=k\text{}^{C}\tau \text{}_{CAB}+k\text{}^{C}\tau \text{}_{CBA}=k\text{}^{C}\tau \text{}_{ABC}=0$ \\
\rownumber & $r_1=r_3=r_4=r_5=t_2=t_3=\lambda =0$ & $\sigma \text{}_{AB}+2 i k\text{}^{C}\tau \text{}_{CBA}=\sigma \text{}_{AB}+2 i k\text{}^{C}\tau \text{}_{ABC}+2 i k\text{}^{C}\tau \text{}_{CAB}=g\text{}^{BC}\tau \text{}_{ACB}=0$ \\
\rownumber & $r_1-r_3=r_4=t_1=t_3=\lambda =0$ & $k\text{}^{C}\tau \text{}_{CAB}+k\text{}^{C}\tau \text{}_{CBA}=i \sigma \text{}_{AB}-k\text{}^{C}\tau \text{}_{ABC}=0$ \\
\rownumber & $r_1=r_2=r_3=r_4=r_5=t_2=t_3=\lambda =0$ & $\sigma \text{}_{AB}+2 i k\text{}^{C}\tau \text{}_{CBA}=g\text{}^{BC}\tau \text{}_{ACB}=\tau \text{}_{ABC}-\tau \text{}_{ACB}+\tau \text{}_{BCA}=0$ \\
\rownumber & $r_2=r_1-r_3=r_4=2 r_1+r_5=t_2=\lambda =0$ & $\sigma \text{}_{AB}+2 i k\text{}^{C}\tau \text{}_{CBA}=\tau \text{}_{ABC}-\tau \text{}_{ACB}+\tau \text{}_{BCA}=0$ \\
\rownumber & $r_1-r_3=r_4=2 r_1+r_5=t_2=\lambda =0$ & $\sigma \text{}_{AB}+2 i k\text{}^{C}\tau \text{}_{CBA}=\sigma \text{}_{AB}+2 i k\text{}^{C}\tau \text{}_{ABC}+2 i k\text{}^{C}\tau \text{}_{CAB}=0$ \\
\rownumber & $r_2=r_1/3-r_3=r_1/3+r_4=2 r_1/3+r_5=t_2=t_3=0$ & $k\text{}^{B}\sigma \text{}_{AB}=\sigma \text{}_{AB}-\sigma \text{}_{BA}-2 i k\text{}^{C}\tau \text{}_{CAB}+2 i k\text{}^{C}\tau \text{}_{CBA}=g\text{}^{BC}\tau \text{}_{ACB}=\tau \text{}_{ABC}-\tau \text{}_{ACB}+\tau \text{}_{BCA}=0$ \\
\rownumber & $r_1/3-r_3=r_1/3+r_4=2 r_1/3+r_5=t_2=t_3=0$ & $k\text{}^{B}\sigma \text{}_{AB}=\sigma \text{}_{AB}-\sigma \text{}_{BA}-2 i k\text{}^{C}\tau \text{}_{CAB}+2 i k\text{}^{C}\tau \text{}_{CBA}=\sigma \text{}_{AB}-\sigma \text{}_{BA}+2 i k\text{}^{C}\tau \text{}_{ABC}=g\text{}^{BC}\tau \text{}_{ACB}=0$ \\
\rownumber & $r_1-r_3=r_4=t_1=t_3=0$ & $k\text{}^{B}\sigma \text{}_{AB}=\sigma \text{}_{AB}-\sigma \text{}_{BA}+2 i k\text{}^{C}\tau \text{}_{ABC}=k\text{}^{C}\tau \text{}_{CAB}+k\text{}^{C}\tau \text{}_{CBA}=0$ \\
\rownumber & $r_1-r_3=r_4=2 r_1+r_5=t_1=t_3=0$ & $k\text{}^{B}\sigma \text{}_{AB}=\sigma \text{}_{AB}-\sigma \text{}_{BA}-2 i k\text{}^{C}\tau \text{}_{CBA}=\sigma \text{}_{AB}-\sigma \text{}_{BA}+i k\text{}^{C}\tau \text{}_{ABC}+i k\text{}^{C}\tau \text{}_{CAB}=0$ \\
\rownumber & $r_1-r_3=r_4=2 r_1+r_5=t_1=t_2=t_3=0$ & $k\text{}^{B}\sigma \text{}_{AB}=\sigma \text{}_{AB}-\sigma \text{}_{BA}=k\text{}^{C}\tau \text{}_{CBA}=k\text{}^{C}\tau \text{}_{ABC}=0$ \\
\rownumber & $r_2=r_1-r_3=r_4=2 r_1+r_5=t_1=t_2=t_3=0$ & $k\text{}^{B}\sigma \text{}_{AB}=\sigma \text{}_{AB}-\sigma \text{}_{BA}=k\text{}^{C}\tau \text{}_{CBA}=\tau \text{}_{ABC}-\tau \text{}_{ACB}+\tau \text{}_{BCA}=0$ \\
\rownumber & $r_1-r_3=r_4=t_1=t_2=t_3=0$ & $k\text{}^{B}\sigma \text{}_{AB}=\sigma \text{}_{AB}-\sigma \text{}_{BA}=k\text{}^{C}\tau \text{}_{CAB}+k\text{}^{C}\tau \text{}_{CBA}=k\text{}^{C}\tau \text{}_{ABC}=0$ \\
\rownumber & $r_2=2 r_3+r_5=t_1=t_2=t_3=\lambda =0$ & $\sigma \text{}_{AB}=\tau \text{}_{ABC}-\tau \text{}_{ACB}+\tau \text{}_{BCA}=k\text{}^{C}\tau \text{}_{CAB}-k\text{}^{C}\tau \text{}_{CBA}=0$ \\
\rownumber & $t_1=t_3=0$ & $k\text{}^{B}\sigma \text{}_{AB}=\sigma \text{}_{AB}-\sigma \text{}_{BA}+2 i k\text{}^{C}\tau \text{}_{ABC}=k\text{}^{B}\sigma \text{}_{BA}=0$ \\
\rownumber & $2 r_3+r_5=t_1=t_2=t_3=\lambda =0$ & $\sigma \text{}_{AB}=k\text{}^{C}\tau \text{}_{CAB}-k\text{}^{C}\tau \text{}_{CBA}=k\text{}^{C}\tau \text{}_{ABC}=0$ \\
\rownumber & $r_2=2 r_3+r_5=t_1=t_2=t_3=0$ & $k\text{}^{B}\sigma \text{}_{AB}=\sigma \text{}_{AB}-\sigma \text{}_{BA}=\tau \text{}_{ABC}-\tau \text{}_{ACB}+\tau \text{}_{BCA}=k\text{}^{C}\tau \text{}_{CAB}-k\text{}^{C}\tau \text{}_{CBA}=0$ \\
\rownumber & $2 r_3+r_5=t_1=t_3=\lambda =0$ & $3 i \sigma \text{}_{AB}-k\text{}^{C}\tau \text{}_{ABC}-k\text{}^{C}\tau \text{}_{CAB}+k\text{}^{C}\tau \text{}_{CBA}=2 i \sigma \text{}_{AB}-k\text{}^{C}\tau \text{}_{CAB}+k\text{}^{C}\tau \text{}_{CBA}=0$ \\
\rownumber & $2 r_3+r_5=t_1=t_3=0$ & $k\text{}^{B}\sigma \text{}_{AB}=\sigma \text{}_{AB}-\sigma \text{}_{BA}+i k\text{}^{C}\tau \text{}_{CAB}-i k\text{}^{C}\tau \text{}_{CBA}=\sigma \text{}_{AB}-\sigma \text{}_{BA}+2 i k\text{}^{C}\tau \text{}_{ABC}=0$ \\
\rownumber & $2 r_3+r_5=t_1=t_2=t_3=0$ & $k\text{}^{B}\sigma \text{}_{AB}=\sigma \text{}_{AB}-\sigma \text{}_{BA}=k\text{}^{C}\tau \text{}_{CAB}-k\text{}^{C}\tau \text{}_{CBA}=k\text{}^{C}\tau \text{}_{ABC}=0$ \\
\rownumber & $t_1=t_2=t_3=0$ & $k\text{}^{B}\sigma \text{}_{AB}=\sigma \text{}_{AB}-\sigma \text{}_{BA}=k\text{}^{C}\tau \text{}_{ABC}=0$ \\
\rownumber & $t_1=t_2=t_3=\lambda =0$ & $\sigma \text{}_{AB}=k\text{}^{C}\tau \text{}_{ABC}=0$ \\
\rownumber & $t_1=t_3=\lambda =0$ & $k\text{}^{B}\sigma \text{}_{AB}=i \sigma \text{}_{AB}-k\text{}^{C}\tau \text{}_{ABC}=0$ \\
\rownumber & $r_2=2 r_3+r_5=t_2=0$ & $k\text{}^{B}\sigma \text{}_{AB}=\sigma \text{}_{AB}-\sigma \text{}_{BA}-2 i k\text{}^{C}\tau \text{}_{CAB}+2 i k\text{}^{C}\tau \text{}_{CBA}=\tau \text{}_{ABC}-\tau \text{}_{ACB}+\tau \text{}_{BCA}=0$ \\
\rownumber & $2 r_3+r_5=t_2=0$ & $k\text{}^{B}\sigma \text{}_{AB}=\sigma \text{}_{AB}-\sigma \text{}_{BA}-2 i k\text{}^{C}\tau \text{}_{CAB}+2 i k\text{}^{C}\tau \text{}_{CBA}=i \sigma \text{}_{AB}-i \sigma \text{}_{BA}-2 k\text{}^{C}\tau \text{}_{ABC}=0$ \\

\end{longtable*}
\egroup
\setcounter{magicrownumbers}{0}

\subsection{Power-counting renormalizability}

\bgroup \def\arraystretch{1.5}
\setlength\tabcolsep{0.11cm}
\begin{table*}[!t]
	\caption{PC renormalizable critical cases that are ghost and
          tachyon free and have only massless propagating
          modes. ``Additional condition'' are the conditions that prevent the
          theory becoming a different critical case. The
          column ``$b$ sectors'' describes the elements in the
          $b^{-1}$-matrix of each spin-parity sector in the sequence
          $(0^-,0^+,1^-,1^+,2^-,2^+)$. Here and 
          in \Cref{tab:PGTUnitaryAndPCNonProp} it is notated as $\varphi^n_{v}$
          or $\varphi^n_{l}$, where $\varphi$ is the field, $-n$ is the
          power of $k$ in the element in the $b^{-1}$-matrix, $v$ means
          massive mode, and $l$ means massless mode.}
	\label{tab:PGTUnitaryAndPC}
	\begin{tabularx}{\textwidth}{lllccc}
		\hline\hline
		\scalebox{.95}[1.0]{No.}&\scalebox{.95}[1.0]{Critical condition}&\scalebox{.95}[1.0]{Additional condition}&\scalebox{.95}[1.0]{No-ghost-and-tachyon condition}&\scalebox{.95}[1.0]{Massless mode d.o.f.}&\scalebox{.9}[1.0]{$b$ sectors}\\ \hline
		\rownumber & \makecell[cl]{$r_3=r_1,r_2=r_4=$\\ $t_1=t_2=t_3=\lambda= 0$}& \makecell[cl]{$r_1\neq 0, r_1+r_5\neq 0$,\\ $2 r_1+r_5\neq 0$} & $r_1 \left(r_1+r_5\right) \left(2 r_1+r_5\right)<0$ & 2 & $(\times,\times,A_l^2,A_l^2,A_l^2,\times)$ \\
		\rownumber & \makecell[cl]{$r_4 = -2 r_1 + 2 r_3,r_2= $\\ $t_1=t_2=t_3=\lambda= 0$}& \makecell[cl]{$r_3\neq 0, 2 r_3+r_5\neq 0 $,\\ $r_3+2 r_5\neq 0,r_2\neq 0$} & $r_1 \left(r_1-2 r_3-r_5\right) \left(2 r_3+r_5\right)>0$ & 2 & $(\times,A_l^2,A_l^2,A_l^2,A_l^2,\times)$ \\
		\rownumber & \makecell[cl]{$r_4=r_3/2, r_1=r_2=$\\ $t_1=t_2=t_3=\lambda= 0$}& \makecell[cl]{$r_3\neq 0, 2 r_3+r_5\neq 0 $,\\ $r_3+2 r_5\neq 0$} & $r_3 \left(2 r_3+r_5\right) \left(r_3+2 r_5\right)<0$ & 2 & $(\times,\times,A_l^2,A_l^2,\times,A_l^2)$ \\
		\rownumber & \makecell[cl]{$r_4=r_3/2, r_1=$\\ $t_1=t_2=t_3=\lambda= 0$}& \makecell[cl]{$r_3\neq 0, 2 r_3+r_5\neq 0 $,\\ $r_3+2 r_5\neq 0,r_2\neq 0$} & $r_3 \left(2 r_3+r_5\right) \left(r_3+2 r_5\right)<0$ & 2 & $(A_l^2,\times,A_l^2,A_l^2,\times,A_l^2)$ \\
		\hline\hline
	\end{tabularx}
\end{table*}
\egroup
\bgroup
\def\arraystretch{1.5}
\setlength\tabcolsep{0.09cm}
\begin{table*}[!t]
	\caption{PC renormalizable critical cases that are ghost and
          tachyon free and have only massive propagating
          modes. The ``$|$''
          notation denotes the different form of the elements of the
          $b^{-1}$-matrices in different choices of gauge fixing.
          The other columns are
          the same as in \Cref{tab:PGTUnitaryAndPC}. Note that while
          there are some $A^0$, $s^2_l$ or $a^2_l$ in the
          $b^{-1}$-matrices, the $A^0$ terms are not propagating. The
          $s^2_l$ or $a^2_l$ terms may lead to PC
          nonrenormalizability, but after applying the source
          constraints and summing all terms from all spin-parity
          sectors, none of the theories below has a massless propagating
          mode. Hence, these terms do not affect PC
          renormalizability.}
	\label{tab:PGTUnitaryAndPCNonProp}
	\begin{tabularx}{\textwidth}{lllccc}
		\hline\hline
		\scalebox{.9}[1.0]{No.}&\scalebox{.9}[1.0]{Critical condition}&\scalebox{.9}[1.0]{Additional condition}&\scalebox{.9}[1.0]{No-ghost-and-tachyon condition}&\scalebox{.9}[1.0]{Massive mode}&\scalebox{.9}[1.0]{$b$ sectors}\\ \hline
		\rownumber &
                \makecell[cl]{$r_1=r_3=r_4=$\\$r_5=t_1=\lambda=0$}&
                \makecell[cl]{$r_2\neq 0,t_2\neq 0,t_3\neq 0$} &
                $t_2>0,r_2<0$  & $0^-$ &
                \makecell[cc]{$(A_v^2,A^0|s^2_l,A^0|s^2_l|a^2_l,A^0|a^2_l,$\\$\times,\times)$}\\
		\rownumber & \makecell[cl]{$r_1=r_3=r_4=$\\$r_5=t_1=t_3=\lambda=0$}& \makecell[cl]{$r_2\neq 0,t_2\neq 0$} & $t_2>0,r_2<0$ & $0^-$ & $(A_v^2,\times,\times,A^0|a^2_l,\times,\times)$\\
		\rownumber & \makecell[cl]{$r_3=r_1,r_5=-2 r_1$,\\$r_4=t_1=t_3= \lambda=0$}& \makecell[cl]{$r_1\neq 0,r_2\neq 0,t_2\neq 0$} & $t_2>0,r_2<0$ & $0^-$ & $(A_v^2,A^0,\times,A^0|a^2_l,\times,\times)$\\
		\rownumber & \makecell[cl]{$r_4=2 r_3,r_5=-2 r_3$, \\$r_1=t_1=t_3=\lambda=0$}& \makecell[cl]{$r_2\neq 0,r_3\neq 0,t_2\neq 0$} & $t_2>0,r_2<0$ & $0^-$ & $(A_v^2,\times,A_l^2,A^0,A_l^2,\times)$\\
		\rownumber & \makecell[cl]{$r_4=r_3/2,r_5=-2 r_3$,\\$r_1=t_1=t_3=\lambda=0$}& \makecell[cl]{$r_2\neq 0, r_3\neq 0,t_2\neq 0$} & $t_2>0,r_2<0$ & $0^-$ & $(A_v^2,\times,A_l^2,A^0|a^2_l,\times,A_l^2)$\\
		\rownumber & \makecell[cl]{$r_4=2 r_3-2 r_1,r_5=-2 r_3$,\\$t_1=t_3=\lambda=0$}& \makecell[cl]{$r_1\neq 0,r_2\neq 0$,\\$ r_1-r_3\neq 0,t_2\neq 0$} & $t_2>0,r_2<0$ & $0^-$ & $(A_v^2,A_l^2,A_l^2,A^0|a^2_l,A_l^2,\times)$\\
		\hline\hline
	\end{tabularx}
\end{table*}
\egroup

In addition to possessing no ghosts or tachyons, a healthy physical
theory should also be renormalizable. The first step in assessing
whether this is possible is to determine whether the theory is
power-counting (PC) renormalizable. 

Even this condition can be quite difficult to establish in the general
case in which the propagator for the theory contains terms that mix
different fields, which is the case for PGT$^+$. Nonetheless, in the
decomposition of the propagator using SPOs, there are some critical
cases for which the mixing terms in the $b$-matrices vanish. In these
cases, the physical meaning is much clearer. We therefore focus only
on the PGT$^+$ critical cases that satisfy this property.

In such cases, one can determine the behavior of the saturated
propagator of the $f$ (graviton) and $A$ (tordion) fields when $k^2\rightarrow
\infty$ by studying the corresponding diagonal elements in the
$b$-matrices. If one requires PC renormalizability, the propagator of
the graviton should go as $k^{-4}$ and that of the tordion should go
as $k^{-2}$ when $k^2\rightarrow \infty$ \cite{Sezgin1980}. We found 10 PC renormalizable critical cases without ghosts and
tachyons, of which four have only massless propagating particles (see
\cref{tab:PGTUnitaryAndPC}) and the remaining six have only a massive
propagating mode (see \Cref{tab:PGTUnitaryAndPCNonProp}). 

It is possible to use different gauge fixing so that sometimes a
graviton mode is transformed to a tordion mode and vice-versa. We find
in these PGT$^+$ cases, however, gauge fixing does not affect
renormalizability. The four cases with only massless modes in
\Cref{tab:PGTUnitaryAndPC} all contain 2 massless degrees of
freedom. There is no way to fix the gauge in these cases without
fixing all the graviton degrees of freedom, so they contain only
tordions. Nonetheless, we note that the inverse $b$-matrices for cases
3 and 4 have elements in the $2^+$ sector, and it might therefore be
of interest to investigate their phenomenology further.  The six cases
in \Cref{tab:PGTUnitaryAndPCNonProp} all propagate only a massive
$0^-$ tordion mode and no massless mode, so they are of limited
physical interest.

We also investigated the PGT$^+$ theories with either zero torsion or
zero curvature, discussed in Secs.~\ref{sec:ztpgt} and
\ref{sec:zcpgt} respectively, but found that no cases are both unitary
and PC renormalizable.

\section{Discussion and Conclusions \label{sec:Conclusion}}

We have presented a systematic method for obtaining the
no-ghost-and-tachyon conditions for all critical cases of a
parity-preserving gauge theory of gravity. We have implemented the
method as a computer program and examined the critical cases of
PGT$^+$, as well as of torsionless PGT$^+$ and teleparallel
PGT$^+$. In comparing our results with the literature for the (small)
subset of critical cases that have been analyzed previously, we find
that they are consistent, apart from a few minor differences that most
probably arise from typographical errors in previous works.

Our method does, however, have the shortcoming that it does not yield
the spins or parities of the massless particles, but only their total
number of degrees of freedom (when there is no dipole ghost). Moreover,
in the presence of a dipole ghost, our method can determine only that
the dipole ghost exists, but does not yield the number of degrees of
freedom.

Although not a shortcoming of our method {\em per se}, it is also
difficult to classify the results obtained. In particular, care must
be taken since, for a given ghost and tachyon free critical case, it
is not guaranteed that all of its child critical cases do not contain
ghosts or tachyons. Furthermore, in general, a theory has multiple
child critical theories, and it also has multiple parent theories, so
it is difficult to divide the theories into some categories without
cutting lots of relations between parent and child theories. Our
interactive interface available at \url{http://www.mrao.cam.ac.uk/projects/gtg/pgt/} is intended
to assist in navigating this space of theories.

An alternative method to that presented here is the Hamiltonian
approach, which has recently been used to study the particle spectrum
of parity-violating PGT by Blagojevi\'{c} and Cvetkovi\'{c}
\cite{Blagojevic2018}. Their results can be straightforwardly reduced
to PGT$^+$ by setting all the $\bar{a}$ and $\bar{b}$ to zero in their
paper. This will not cause any new ``critical parameters" to
vanish. By comparing their ``critical parameters" with our ``critical
conditions," we find that our type C critical conditions are identical
to their critical parameters. These critical parameters are second
class constraints \cite{Blagojevic1983,Nikolic}, so they do not lead
to additional gauge invariance, which is consistent with our
definition of type C critical cases. As for the type A critical
conditions, we believe that they correspond to first class
if-constraints because first class constraints represent additional
gauge invariance. In Blagojevic's book
\cite{Blagojevic2002}, the critical parameters for the most general
teleparallel PGT$^+$ are listed, and found to be first class. Our
method found 4 type A conditions from the theory, which is the same as
Blagojevic. This is consistent with our supposition. As for the type B
critical cases, however, \cite{Blagojevic2018} does not
mention its consequences (massive particle becomes massless), but only
requires the mass squares to be positive. Blagojevi\'{c} and
Vasili\'{c} \cite{Blagojevic1987} studied what happens when massive
modes becomes massless. In particular, they claim that if any massive
tordion becomes massless, there will be extra gauge
invariance. However, in their analysis they always include other
critical condition(s) in addition to setting the mass to zero to make
the theory healthy, so they are not purely applying type B
conditions. It is possible that we combine some type B conditions with
some other conditions to get a type A condition and extra gauge
invariance appears, so their conclusion does not conflict with ours.

In the context of PGT$^+$, it may be of interest to investigate
further the theories listed in \Cref{tab:PGTUnitaryAndPC}, which are
both unitary and power-counting renormalizable, and possess only
massless propagating particles.  Although these theories contain no
graviton, only tordions, they may provide some insights into the
construction of a self-consistent quantum theory of long-range
gravitational interactions. In particular, cases 3 and 4 might be of
interest, since they may possess particles in the $2^+$
sector. Indeed, it is worth noting that in the absence of torsion the
action for both of these cases reduces to that of conformal gravity,
which is PC renormalizable but not unitary, as discussed in
Case~\ref{row:NormalConformal} in Sec.~\ref{sec:comparison}. 

Finally, although we demonstrated our method only for PGT$^+$ in this paper, it
may be applied to more complex theories such as Weyl gauge theory
(WGT) \cite{Bregman1973,*Charap1974,*Kasuya1975} or extended Weyl
gauge theory (eWGT) \cite{Lasenby2016}. It is also applicable to
conventional metric theories such as $\mathcal{R}^2$ theories. We plan
to explore its application to such theories in future work. 

\begin{acknowledgments}
Y.-C. Lin acknowledges support from the Ministry of Education of
Taiwan and the Cambridge Commonwealth, European \& International Trust
via a Taiwan Cambridge Scholarship.
\end{acknowledgments}
\medskip

\onecolumngrid
\appendix

\section{SPIN PROJECTION OPERATORS FOR PGT$^+$ \label{sec:SpinProjectionOperator}}

The block matrices $\mathsf{P}(J^P)$ containing the spin projection
operators for PGT$^+$ used in this paper are as follows (see
Sec.~\ref{section:primer} for details):

	\begin{align}
	&\mathsf{P}\left( 0^-\right)=
	\bordermatrix{
		~& A_{ABC} \cr
		A^*_{IJK}& \frac{2}{3} \Theta \text{}_I\text{}_C \Theta \text{}_J\text{}_A \Theta \text{}_K\text{}_B+\frac{1}{3} \Theta \text{}_I\text{}_A \Theta \text{}_J\text{}_B \Theta \text{}_K\text{}_C \cr
	},
	\\[3mm]
	&\mathsf{P}\left( 0^+\right)=
	\bordermatrix{
		~&A_{ABC}&s_{AB}&s_{AB} \cr
		A^*_{IJK}&\frac{2}{3} \Theta \text{}_C\text{}_B \Theta \text{}_K\text{}_J \Omega \text{}_I\text{}_A & \frac{\sqrt{2}}{3} \tilde{k}\text{}_J \Theta \text{}_A\text{}_B \Theta \text{}_K\text{}_I & \sqrt{\frac{2}{3}} \tilde{k}\text{}_J \Theta \text{}_K\text{}_I \Omega \text{}_B\text{}_A  \cr
		s^*_{IJ}&\frac{\sqrt{2}}{3} \tilde{k}\text{}_B \Theta \text{}_C\text{}_A \Theta \text{}_I\text{}_J & \frac{1}{3} \Theta \text{}_A\text{}_B \Theta \text{}_I\text{}_J & \frac{1}{\sqrt{3}}\Theta \text{}_I\text{}_J \Omega \text{}_A\text{}_B  \cr
		s^*_{IJ}&\sqrt{\frac{2}{3}}\tilde{k}\text{}_B \Theta \text{}_C\text{}_A \Omega \text{}_J\text{}_I & \frac{1}{\sqrt{3}}\Theta \text{}_A\text{}_B \Omega \text{}_I\text{}_J & \Omega \text{}_A\text{}_B \Omega \text{}_I\text{}_J  \cr
	},
	\\[3mm]
	&\mathsf{P}\left( 1^-\right)=
	\bordermatrix{
		~&A_{ABC}&A_{ABC}&s_{AB}&a_{AB} \cr
		A^*_{IJK}&\Theta \text{}_C\text{}_B \Theta \text{}_I\text{}_A \Theta \text{}_K\text{}_J & \sqrt{2} \Theta \text{}_I\text{}_A \Theta \text{}_K\text{}_J \Omega \text{}_C\text{}_B & \sqrt{2} \tilde{k}\text{}_B \Theta \text{}_I\text{}_A \Theta \text{}_K\text{}_J & \sqrt{2} \tilde{k}\text{}_B \Theta \text{}_I\text{}_A \Theta \text{}_K\text{}_J  \cr
		A^*_{IJK}&\sqrt{2} \Theta \text{}_A\text{}_I \Theta \text{}_C\text{}_B \Omega \text{}_K\text{}_J & 2 \Theta \text{}_I\text{}_A \Omega \text{}_C\text{}_B \Omega \text{}_K\text{}_J & 2 \tilde{k}\text{}_J \Theta \text{}_I\text{}_A \Omega \text{}_K\text{}_B & 2 \tilde{k}\text{}_J \Theta \text{}_I\text{}_A \Omega \text{}_K\text{}_B  \cr
		s^*_{IJ}&\sqrt{2} \tilde{k}\text{}_J \Theta \text{}_A\text{}_I \Theta \text{}_C\text{}_B & 2 \tilde{k}\text{}_B \Theta \text{}_A\text{}_I \Omega \text{}_C\text{}_J & 2 \Theta \text{}_I\text{}_A \Omega \text{}_J\text{}_B & 2 \Theta \text{}_I\text{}_A \Omega \text{}_J\text{}_B  \cr
		a^*_{IJ}&\sqrt{2} \tilde{k}\text{}_J \Theta \text{}_A\text{}_I \Theta \text{}_C\text{}_B & 2 \tilde{k}\text{}_B \Theta \text{}_I\text{}_A \Omega \text{}_C\text{}_J & 2 \Theta \text{}_I\text{}_A \Omega \text{}_J\text{}_B & 2 \Theta \text{}_I\text{}_A \Omega \text{}_J\text{}_B  \cr
	},
	\\[3mm]
	&\mathsf{P}\left( 1^+\right)=
	\bordermatrix{
		~&A_{ABC}&A_{ABC}&a_{AB} \cr
		A^*_{IJK}&\Theta \text{}_I\text{}_C \Theta \text{}_K\text{}_B \Omega \text{}_J\text{}_A+\Theta \text{}_I\text{}_A \Theta \text{}_K\text{}_C \Omega \text{}_J\text{}_B & -\sqrt{2} \Theta \text{}_J\text{}_A \Theta \text{}_K\text{}_B \Omega \text{}_I\text{}_C & \sqrt{2} \tilde{k}\text{}_J \Theta \text{}_I\text{}_A \Theta \text{}_K\text{}_B \cr
		A^*_{IJK}&-\sqrt{2} \Theta \text{}_B\text{}_I \Theta \text{}_C\text{}_J \Omega \text{}_A\text{}_K & \Theta \text{}_I\text{}_A \Theta \text{}_J\text{}_B \Omega \text{}_K\text{}_C & \tilde{k}\text{}_K \Theta \text{}_I\text{}_A \Theta \text{}_J\text{}_B \cr
		a^*_{IJ}&\sqrt{2} \tilde{k}\text{}_B \Theta \text{}_A\text{}_I \Theta \text{}_C\text{}_J & \tilde{k}\text{}_C \Theta \text{}_A\text{}_I \Theta \text{}_B\text{}_J & \Theta \text{}_A\text{}_I \Theta \text{}_B\text{}_J \cr
	},
	\\[3mm]
	&\mathsf{P}\left( 2^-\right)=
	\bordermatrix{
		~&A_{ABC} \cr
		A^*_{IJK}&\frac{2}{3} \Theta \text{}_I\text{}_C \Theta \text{}_J\text{}_B \Theta \text{}_K\text{}_A+\frac{2}{3} \Theta \text{}_I\text{}_A \Theta \text{}_J\text{}_B \Theta \text{}_K\text{}_C-\Theta \text{}_C\text{}_B \Theta \text{}_I\text{}_A \Theta \text{}_K\text{}_J \cr
	}, \\[3mm]
	&\mathsf{P}\left(2^+\right)=
	\bordermatrix{
		~&A_{ABC}&s_{AB} \cr
		A^*_{IJK}&-\frac{2}{3} \Theta \text{}_C\text{}_B \Theta \text{}_K\text{}_J \Omega \text{}_I\text{}_A+\Theta \text{}_I\text{}_C \Theta \text{}_K\text{}_A \Omega \text{}_J\text{}_B+\Theta \text{}_I\text{}_A \Theta \text{}_K\text{}_C \Omega \text{}_J\text{}_B & \sqrt{2} \tilde{k}\text{}_J \left(\Theta \text{}_I\text{}_A \Theta \text{}_K\text{}_B-\frac{1}{3} \Theta \text{}_A\text{}_B \Theta \text{}_K\text{}_I\right) \cr
		s^*_{IJ}&\sqrt{2} \tilde{k}\text{}_B \left(\Theta \text{}_C\text{}_J \Theta \text{}_I\text{}_A-\frac{1}{3} \Theta \text{}_C\text{}_A \Theta \text{}_I\text{}_J\right) & -\frac{1}{3} \Theta \text{}_A\text{}_B \Theta \text{}_I\text{}_J+\Theta \text{}_I\text{}_A \Theta \text{}_J\text{}_B \cr
	},\\\nonumber
	\end{align}
where $\tilde{k}\text{}_A=k\text{}_A/\sqrt{k^2}$, $\Omega^{AB}=k^{A}
k^{B}/k^2$, and $\Theta^{AB}=\eta^{AB}-k^{A} k^{B}/k^2$.  The
operators are adapted from \cite{Karananas2015}. The fields
have some symmetry properties: the $A_{ABC}$ field is antisymmetric in
$AB$, the $a_{AB}$ field is antisymmetric in $AB$, and the $s_{AB}$
field is symmetric in $AB$. Note that the spin projection operators
satisfy the symmetry properties implicitly. For example, although
$P_{33}(1^-) = P^{(ss)}_{11}(1^-)$ is notated as
$2\Theta_{IA}\Omega_{JB}$ above, its correctly symmetrized form is
$\para{\Theta_{IA}\Omega_{JB}+\Theta_{IB}\Omega_{JA}+\Theta_{JA}\Omega_{IB}+\Theta_{JB}\Omega_{IA}}/2$. We
have verified that the above set of spin projection operators
satisfies \eqref{eqn:Pcomplete} and \eqref{eqn:Portho}.

\twocolumngrid

\bibliography{GaugeGravity_abbr}

\end{document}
%